\documentclass[twocolumn]{aastex631}
\usepackage{CJKutf8}

\submitjournal{ApJ on June 24th 2024, Accepted September 3rd 2024}
\shorttitle{YSG Binaries: Identification}
\shortauthors{O'Grady et al.}

\defcitealias{Neugent.K.2018.RSGBinaryMethod}{N18}

\begin{document}
\begin{CJK*}{UTF8}{gbsn}

\title{Binary Yellow Supergiants in the Magellanic Clouds I: Photometric Candidate Identification}
\correspondingauthor{Anna O'Grady}
\email{aogrady@andrew.cmu.edu}

\author[0000-0002-7296-6547]{Anna J. G. O'Grady}
\affil{McWilliams Center for Cosmology \& Astrophysics, Department of Physics, Carnegie Mellon University, Pittsburgh, PA 15213, USA}
\affil{David A. Dunlap Department of Astronomy and Astrophysics, University of Toronto, 50 St. George Street, Toronto, ON, M5S3H4 Canada}
\affil{Dunlap Institute for Astronomy and Astrophysics, University of Toronto, 50 St. George Street, Toronto, ON, M5S3H4, Canada}

\author[0000-0001-7081-0082]{Maria R. Drout}
\affil{David A. Dunlap Department of Astronomy and Astrophysics, University of Toronto, 50 St. George Street, Toronto, ON, M5S3H4 Canada}

\author[0000-0002-5787-138X]{Kathryn F. Neugent}
\affil{Center for Astrophysics \text{\textbar} Harvard \& Smithsonian, 60 Garden Street, Cambridge, MA 02138-1516, USA}

\author[0000-0003-0857-2989]{Bethany Ludwig}
\affil{David A. Dunlap Department of Astronomy and Astrophysics, University of Toronto, 50 St. George Street, Toronto, ON, M5S3H4 Canada}
\affil{Dunlap Institute for Astronomy and Astrophysics, University of Toronto, 50 St. George Street, Toronto, ON, M5S3H4, Canada}

\author[0000-0002-6960-6911]{Ylva G\"{o}tberg}
\affil{Institute of Science and Technology Austria (ISTA), Am Campus 1, 3400 Klosterneuburg, Austria}
\affil{The Observatories of the Carnegie Institution for Science, 813 Santa Barbara St., Pasadena, CA 91101, USA}

\author[0000-0002-3382-9558]{B. M. Gaensler}
\affil{Dunlap Institute for Astronomy and Astrophysics, University of Toronto, 50 St. George Street, Toronto, ON, M5S3H4, Canada}
\affil{David A. Dunlap Department of Astronomy and Astrophysics, University of Toronto, 50 St. George Street, Toronto, ON, M5S3H4 Canada}
\affil{Department of Astronomy and Astrophysics, University of California Santa Cruz, 1156 High Street, Santa Cruz, CA 95064, USA}

\begin{abstract}

Recent works have constrained the binary fraction of evolved populations of massive stars in local galaxies such as red supergiants and Wolf-Rayet stars, but the binary fraction of yellow supergiants (YSGs) in the Hertzsprung Gap remains unconstrained. Binary evolution theory predicts that the Hertzsprung Gap is home to multiple populations of binary systems with varied evolutionary histories. In this paper, we develop a method to distinguish single YSGs from YSG plus O- or B-type main sequence binaries using optical and ultraviolet photometry, and then apply this method to identify candidate YSG binaries in the Magellanic Clouds. After constructing a set of combined stellar atmosphere models, we find that optical photometry is, given typical measurement and reddening uncertainties, sufficient to discern single YSGs from YSG+OB binaries if the OB-star is at least $\sim5$M$_{\odot}$ for T$_{\mathrm{eff,YSG}}\sim$ 4000 K, but requires a $\sim$20M$_{\odot}$ OB star for YSGs up to T$_{\mathrm{eff,YSG}}\sim$ 9000 K. For these hotter YSG temperatures, ultraviolet photometry allows binaries with OB companions as small as $\sim$7M$_{\odot}$ to be identified. We use color-color spaces developed from these models to search for evidence of excess blue or ultraviolet light in a set of $\sim$1000 YSG candidates in the Magellanic Clouds. We identify hundreds of candidate YSG binary systems and report a preliminary fraction of YSGs that show a blue/UV color excess of 20-60\%. Spectroscopic follow-up is now required to confirm the true nature of this population. 

\end{abstract}

\section{Introduction}\label{sec:intro}

The importance and ubiquity of binarity to the evolution of massive stars has become increasingly clear over the past decade. Much work has been done to constrain the binary fraction of main sequence (MS) stars; we now know that 70-100\% of massive (M$\gtrapprox$8M$_{\odot}$) stars on the MS have binary companions \citep{Roberts.L.2007.BStarMulti,Chini.R.2012.BinaryFraction,Kiminki.D.2012.BinaryFractionO,Sana.H.2012.BinaryFraction,Sana.H.2014.SouthernStarMultiplicity,Rizzuto.A.2013.BStarMulti,Kobulnicky.H.2014.BinariesOB2,Banyard.G.2022.BStarMulti,Offner.S.2023.MultipleStarSystems}. This has strong implications for the subsequent evolution of massive binaries. When coupled with measurements of the period and mass ratio distribution \citep{Moe.M.2017.BinaryPeriodMass}, binary population synthesis codes predict that $\sim$70\% of massive binaries with \emph{interact} at some point in their evolution \citep{Sana.H.2012.BinaryFraction}. However, there remain many uncertainties in binary population models for how interactions are modeled, from the efficiency of binary mass transfer to outcomes of common envelope evolution and supernova kicks \citep{deKool.M.1990.CEE1,Soberman.G.1997.MassTransferStability,Belczynski.K.2007.CEE3,Ivanova.N.2013.CEE2,Schneider.F.2015.MassTransferEff,Bray.J.2016.Kicks2,Giacobbo.N.2020.Kicks1}. These uncertainties have broad consequences on our understanding of binary evolution processes; for example, rates of compact object mergers (and thus gravitational wave predictions) are uncertain to several orders of magnitude \citep{Broekgaarden.F.2022.CompactMergerRates,Mandel.I.2022.CompactObjectMergers}. It is therefore vital to supplement measurements of the binary fractions obtained during the MS phase with constraints and empirical rates from evolved stars and post-interaction binaries.

Several studies have put observational constraints on the occurrence rates and properties of massive evolved binary systems in both the Milky Way and nearby galaxies. \citet{Neugent.K.2014.WRBinFrac} found an apparent binary fraction of 30\% for Wolf-Rayet stars in M31 and M33. For red supergiants (RSGs), \citet{Patrick.L.2019.RSGBin30Dor,Patrick.L.2020.RSGBinNGC330} found a binary fraction of $\sim$30\% when investigating clusters, \citet{Neugent.K.2020.RSGBinaryLMC} found a binary fraction of 20\% in the Large Magellanic Cloud, and \citet{Neugent.K.2021.RSGBinM3133} found a binary fraction of 34\% in M33 and a range of $\sim$16-41\% from low- to high-metallicity regions in M31. As well, hot subdwarf companions of Galactic Be stars have been identified \citep{Wang.L.2017.HotSD1,Wang.L.2018.HotSD2,Wang.L.2021.HotSD3}, as well as Wolf-Rayet+O binaries \citep{Shara.M.2017.WROBin,Shara.M.2020.WROBin2,Shenar.T.2019.WROBin3}. There are also ongoing efforts to search for the products of the predicted 20--30\% of massive binaries that will merge on the main sequence \citep{Blagorodnova.N.2017.YSGasCEEjection,Blagorodnova.N.2021.YSGLRNebwo,Schneider.F.2019.MergersMagStars}. The later evolutionary stages of detached massive binaries such as intermediate mass stripped stars \citep{DroutGotberg.2023.IntStrippedStars}, X-ray MS+ black hole systems \citep{Liu.J.2019.MassiveBinBlackHole3,Rivinius.T.2020.MassiveBinBlackHole2,Sen.K.2021.MassiveBinBlackHole4,Shenar.T.2022.MassiveBinBlackHole}, and massive X-ray binaries \citep{Liu.Q.2006.HMXBGalaxy,Haberl.F.2016.HMXBinSMC,Antoniou.V.2016.HMXBinLMC,Lazzarini.M.2021.HMXBM31}, have also been investigated.

One important phase of massive star evolution for which there are not yet estimates of the binary fraction is the yellow supergiant (YSG) phase. In the traditional picture of single-star evolution, YSGs represent a short ($\sim$ 10,000 yrs) transitional phase in which evolved massive stars with $\sim8<$ M/M$_{\odot}<30$ move through the cooler portion of the Hertzsprung Gap, their temperatures cooling and radii inflating dramatically, until they reach the RSG branch where they are expected to explode as hydrogen-rich supernovae \citep[e.g.,][]{Ekstrom.S.2012.Models1Rotation}. However, the presence of a binary companion can strongly alter the evolutionary pathway that a massive star takes through the Hertzsprung-Russell diagram. While 20-30\% of massive binaries have short orbital periods and are predicted to merge on the MS, systems with initial orbital periods in the range of $\sim$10--3000 days are expected to interact at later evolutionary stages as the primary star evolves and dramatically increases in radius during its crossing of the Hertzsprung-Russell diagram \citep{Sana.H.2012.BinaryFraction}. Thus, the Hertzsprung Gap is expected to be the site of many post-MS binary interactions, the precise physical details of whose mechanisms are not well constrained.

As well, pre-explosion imaging of several supernovae (SNe) has indicated that the Hertzsprung Gap plays an important role in the terminal phase for some massive stars. In particular, the progenitors of some partially stripped (Type IIb) supernovae have been identified as YSGs \citep{Aldering.G.1994.SN93JProgen,Crockett.R.2008ax,Maund.J.2011.SN2011dhProgenitor,Eldridge.J.2013.IIbYSGPlot,VanDyk.S.2014.2013dfYSG}. One proposed explanation is that enhanced mass-loss during the RSG phase could cause some stars to transition back to hotter temperatures and then explode in the Hertzsprung Gap \citep{Georgy.C.2012.MassLossRSGBlueward}, but the physical mechanism that would lead to this is unknown, and there is debate over whether the currently-adopted mass-loss rates of RSGs are over- or under-estimated \citep[e.g.][]{Beasor.E.2020.RSGMLR,Massey.P.2023.RSGMLR}. Another proposed explanation is binary interaction. For some binary separations and mass ratios, binary interaction will lead to partial envelope stripping, leaving the primary with a thin hydrogen envelope \citep{Yoon.S.2010.YSGIIbBinary,Yoon.Sung-Chul.2017.TypeIbIIbBinarity}. Current binary evolution models are able to reproduce both the ejecta masses and pre-explosion mass loss rates inferred from supernova observations \citep{Yoon.S.2010.YSGIIbBinary,Sravan.N.2020.IIbProgenitorsObs,Laplace.E.2020.SEStars}. As well, there are tentative detections of massive, blue binary companions at the precise location of previous partially stripped-envelope SNe \citep{Maund.J.2004.1993JProgenitor,Folatelli.G.2014.SN2011dhBluePoint,Ryder.S.2018.SN2001igBinary}. 

A robust understanding of how binary companions can affect the YSG phase is thus crucial, both for understanding post-MS binary interaction and the progenitors of partially stripped-envelope SNe. However, while significant progress has been made in systematically identifying and characterizing YSGs in local galaxies \citep{Drout.M.2009.YSGinM31,Drout.Maria.2012.RSG.YSG.M33,Neugent.K.2010.YSGinSMC,Neugent.K.2012.RSG.YSG.LMC}, there have been few observational constraints put on populations of YSG binaries. There have been searches for hot companions of luminous yellow stars at lower mass ranges \citep{ArellanoFerro.A.1986.UVSearchHotYSGs,Evans.N.1992.CepheidBinarySearch}, and a handful of individual YSG binary systems have been suggested or identified \citep{ArellanoFerro.A.1981.VarYSGsinMWG,ArellanoFerro.A.1984.89HercBin,Sperauskas.J.2014.YSGBinSpecEx}. There are also some observed examples of YSG binaries in evolved interaction states, including systems undergoing RLOF \citep{Prieto.J.2008.YSGEBRLOF}, undergoing common envelope evolution \citep{Chesneau.O.2014.YSGinCEE}, or as the progenitor of a luminous red novae common envelope ejection event \citep{Blagorodnova.N.2017.YSGasCEEjection,Blagorodnova.N.2021.YSGLRNebwo,Cai.Y.2022.YSGLRNeHugs}. But there has been no systematic identification of YSG binary systems, nor constraint on the YSG binary fraction. In order to tackle these challenges, we must first understand what types of YSG binary systems are detectable via different methods.

To that end, we aim to develop a method to systematically identify candidate YSG binary systems using optical and ultraviolet (UV) photometry, with further goals of characterizing YSG binaries and constraining the YSG binary fraction in the Magellanic Clouds. In this paper, we use stellar atmosphere models of YSGs and their likely companions to identify regions of color-color space that YSG binaries inhabit, allowing for photometric classification of probable YSG binary systems. We apply this method to candidate YSGs in the Magellanic Clouds and find success in identifying hundreds of candidate binary systems.
In \S\ref{sec:method} we describe our method of photometrically identifying YSG binary systems, and in \S\ref{sec:results} we report the results of applying this method to a population of YSG candidates in the Magellanic Clouds. In \S\ref{sec:disc} we discuss the robustness of this method and implications of our results. Finally in \S\ref{sec:concl} we conclude and discuss future directions.

\section{Methodology for Photometrically Identifying YSG Binaries}\label{sec:method}

As there are few observed examples of binary YSGs, we combine synthetic models of YSGs with models of their likely companions to define a set of photometric selection criteria that can be used to distinguish between single YSGs and YSG binary systems. Here we describe the components and steps of the methodology used to identify these YSG binary candidates. We generally follow the method used by \citet[][hereafter \citetalias{Neugent.K.2018.RSGBinaryMethod}]{Neugent.K.2018.RSGBinaryMethod} to identify candidate RSG binaries. 

\subsection{Expected Components of YSG Binary Systems}\label{sec:components}

\subsubsection{Definition of a Yellow Supergiant}\label{sec:ysgdef}

In this work, we broadly consider stars to be YSGs if they are located within the cooler portion of Hertzsprung Gap (but warmer than the Hayashi line) and are massive enough to explode as supernovae. Specifically, we define YSGs to be stars with log(L/L$_{\odot}$) $\geq$ 4.0 and effective temperature 4000 $\leq$ T$_{\mathrm{eff}}$ $\leq$ 9000 K. These temperatures correspond to spectral classes A-F-G-K, and the luminosity cut corresponds to stars with M $\gtrsim$ 9M$_{\odot}$ using the Geneva stellar evolutionary models \citep{Lejeune.T.2001.GenevaPhotModels}. When defining this temperature cut, we broadly follow the works of \citet{Drout.M.2009.YSGinM31,Drout.Maria.2012.RSG.YSG.M33} and \citet{Neugent.K.2010.YSGinSMC,Neugent.K.2012.RSG.YSG.LMC}. However, we chose wider T$_{\mathrm{eff}}$ bounds to encompass known YSG SNe progenitors---which span $\sim$4200 K to 8900 K for SN1993J and SN2008ax, respectively (\citealt{Maund.J.2004.1993JProgenitor,Crockett.R.2008ax}; see also Fig 12, Fig 6 of \citealt{Eldridge.J.2013.IIbYSGPlot,Eldridge.J.2018.SNPopBinaries}). The lower end of our temperature range also avoids the RSG branch, which is located around $\sim$4000/3800 K in the SMC/LMC, respectively \citep{Ekstrom.S.2012.Models1Rotation}.

\subsubsection{The Companions of YSGs in Binaries}\label{sec:companionsofysgs}

As described in \S\ref{sec:intro}, there are multiple evolutionary pathways that can lead to a YSG with a binary companion. These include: i) a wide binary crossing the Hertzsprung Gap for the first time before becoming a RSG, or ii) an interacting or post-interaction binary. 

In the first case, these wide binaries are the progenitors of RSG binary systems, such as those identified by \citet{Neugent.K.2019.RSGBinaryDiscovering,Neugent.K.2020.RSGBinaryLMC}, \citet{Neugent.K.2021.RSGBinM3133}, and \citet{Patrick.L.2017.RSGBinNGC55,Patrick.L.2019.RSGBin30Dor,Patrick.L.2020.RSGBinNGC330}. These are typically RSG+B-type star systems; \citetalias{Neugent.K.2018.RSGBinaryMethod} demonstrated through evolutionary models that for binaries that haven't interacted upon the primary reaching the RSG branch, B-type stars are the most likely type of companion. Specifically, companion stars with M$<$3M$_{\odot}$ will not have reached the zero-age main sequence (ZAMS) by the time the primary begins ascending the RSG branch, and O-type stars (M$\gtrapprox$16M$_{\odot}$) are much rarer than B-type stars. We therefore expect the YSG binary systems that have not undergone interactions to also predominantly have B-type star companions.

The second case requires a binary system that does not merge on the MS, but interacts later in life. Within this context, there are multiple phases of evolution when one of the components of an interacting/post-interaction binary may appear in the cooler portion of the Hertzsprung gap. For example:

\begin{itemize}
    \item \emph{Currently or recently interacting systems:} For certain initial period separations, a binary system will interact when the primary star enters the YSG regime. For less extreme mass ratios, the primary star can remain in the YSG regime while undergoing a short (thermal timescale) phase of stable mass transfer. Examples of current or post-mass transfer systems in the hot end of the Herzsprung Gap have been identified \citep{Ramachandran.V.2023.PartiallyStrippedBinaryEx,Villasenor.J.2023.BTypeStrippedStar}.
    \item \emph{Post helium burning envelope expansion:} While (partially) stripped stars are expected to be hot and compact for most of their lives \citep{2018A&A...615A..78G}, after helium core-burning ceases they will undergo envelope expansion. Depending on metallicity and hydrogen envelope mass, they can (re)enter the YSG regime \citep{Laplace.E.2020.SEStars}.
    \item \emph{Helium burning at low metallicity: } \citet{Klencki.J.2022.PartialEnvelopeMissingStars} recently showed that in low metallicity environments, some partially stripped stars can remain at cool temperatures even during the core helium-burning phase, leading to nuclear timescale mass transfer and detached inflated systems.
\end{itemize}
\noindent In cases where the YSG is the primary star, the lower mass secondary likely was a B-type star at ZAMS by the same arguments given above for non-interacting YSG binaries. However, if the system undergoes a phase of conservative mass transfer, the secondary could potentially gain significant mass. Indeed, in a population study, \citet{Zapartas.E.2017.SESNeCompanions} predicted that $\sim$68\% of (partially-)stripped-envelope SN progenitors have MS stars as companions (while $\sim$5\% have compact object companions and $\sim$26\% die single). Of these MS companions, they find that $\sim$63\% are 2--16M$_{\odot}$ B-type stars, while 37\% are 16--60 M$_{\odot}$ O-type stars.

From this, we follow \citetalias{Neugent.K.2018.RSGBinaryMethod} in assuming that the companions of non-interacting YSG binaries will be dominated by B-type stars. For interacting systems, mass transfer can lead to a higher fraction of O-type stars.

\subsection{Combined Stellar Atmosphere Models}\label{sec:combinedmodels}

\begin{figure*}
    \centering
    \includegraphics[width=0.9\textwidth]{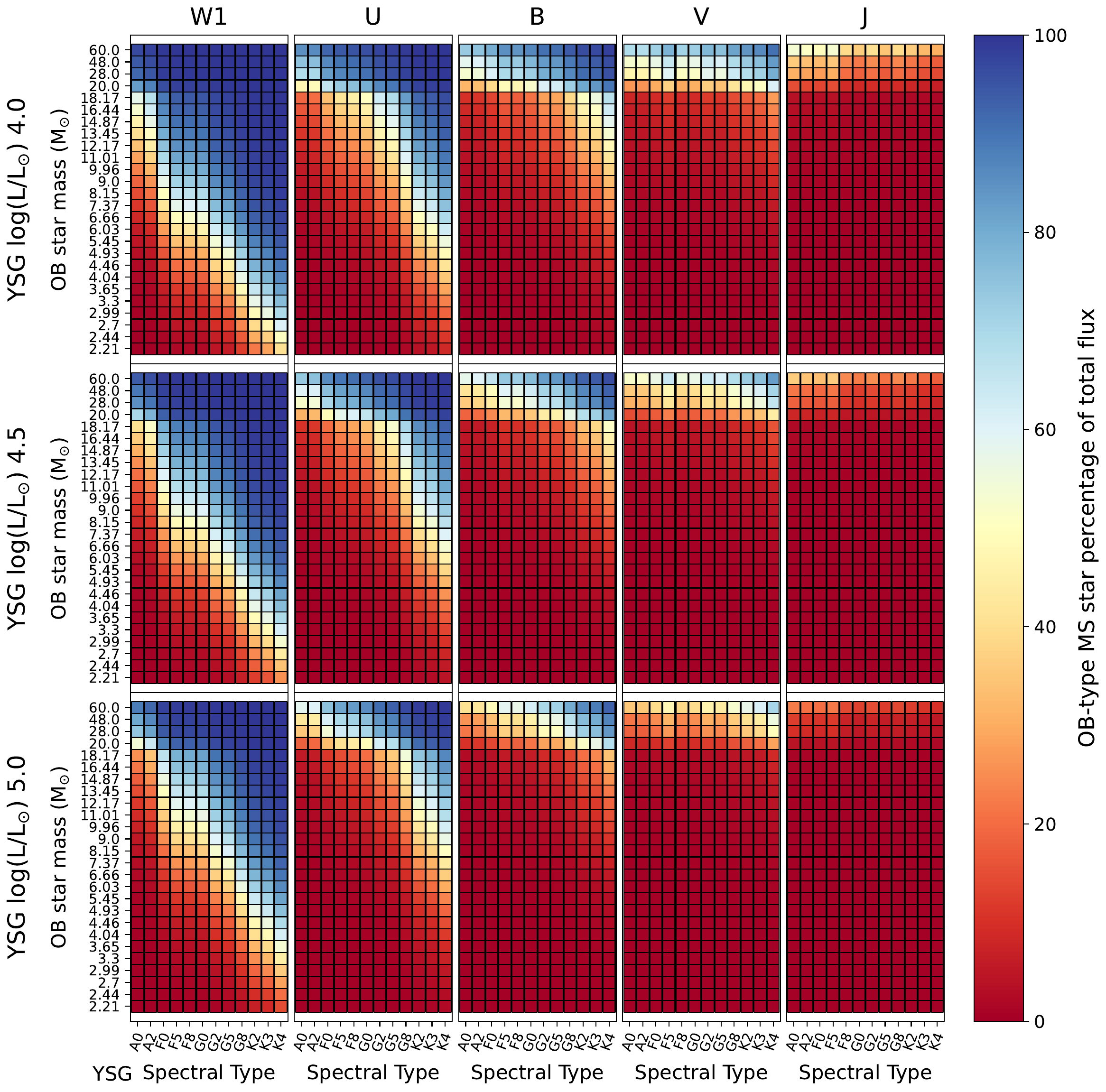}
    \caption{The percentage that the OB-type MS star contributes towards the total flux in the YSG+OB binary system. Each column shows the results for a filter (from left to right, W1 (Swift), U, B, and V (MCPS), and J (2MASS)), and each row shows the results for a given YSG luminosity (from top to bottom, log(L/L$_{\odot}$) = 4.0, 4.5, 5.0). Within the panels, each box represents a combined model for a given YSG spectral type and OB-star mass.}
    \label{fig:fracflux}
\end{figure*}

Based on \S\ref{sec:components}, we combine a set of synthetic YSG models with both O- and B-type (OB) MS models in order to understand how photometric colors from a YSG+OB binary will differ from those of single YSGs. 
For all of the following models, both single and combined, we compute synthetic photometry for filters UV-W1 (referred to simply as W1 from hereafter) from the \emph{Swift} mission \citep{Gehrels.N.2004.Swift,Roming.P.2005.SUVOT}, U, B, V, I from the Magellanic Cloud Photometric Survey \citep[MCPS,][]{Zaritsky.D.2002.MCPSSMC,Zaritsky.D.2004.MCPSLMC}, and J and K$_{\mathrm{s}}$ from the 2 Micron All Sky Survey \citep[2MASS,][]{2MASS.Skrutskie.2006}. Filter profiles were obtained from SVO Filter service\footnote{\url{http://svo2.cab.inta-csic.es/theory/fps/}} \citep{Rodrigo.C.2012.SVOFilterService}.

We use YSG models from the Pickles Stellar Spectral Flux library \footnote{\url{https://vizier.cds.unistra.fr/viz-bin/VizieR?-source=J/PASP/110/863}} \citep{Pickles.A.1998.SynthSpec}. We select 12 luminosity class I models with A, F, G, and K spectral types, which correspond to temperatures of 3900-9700 K, K4-A0; the hottest model, A0, is for demonstrative purposes only. An A2 model at 9000 K serves as the hot boundary model. We then create four copies of each spectral model, which we scale to luminosities of log(L/L$_{\odot}$) = 4.2, 4.5, 4.8, and 5.1.  When compared to Geneva stellar evolutionary models \citep{Lejeune.T.2001.GenevaPhotModels}, these luminosities correspond to YSGs with masses of approximately 9, 12, 15, and 20M$_{\odot}$, respectively. In order to scale the models to a given luminosity, we first determine the expected K-band magnitude for a YSG at a given temperature and luminosity using a J$-$K versus K-band relationship for stars in this portion of the HR diagram (see \S\ref{init_cc} for additional details). We then scaled the spectra of the Pickles models based on the difference between their original (arbitrary) K-band magnitudes and the true K-band magnitudes expected for each luminosity.

For the OB-type companion stars, we use two sets of stellar atmosphere models computed with CMFGEN \citep{1990A&A...231..116H, 1998ApJ...496..407H}. The first set was initially presented by \citet{DroutGotberg.2023.IntStrippedStars}, and includes 22 models in total, spanning ZAMS masses of 2.21--18.17 M$_{\odot}$. The models are described in the supplementary material of \citet{DroutGotberg.2023.IntStrippedStars} and are available online \citep{gotberg_2023_7976200}. In brief: a set of MESA evolutionary models computed by \citet{2018A&A...615A..78G} were used to define the surface properties at the base of the atmosphere. While \citet{DroutGotberg.2023.IntStrippedStars} present models for stars at various points of the MS, for simplicity, we use early MS models that are 20\% of their way through the MS. These models have luminosities of log(L/L$_{\odot}$) = 1.35--4.55, temperatures of T$_{\mathrm{eff}}$ = 12800--35000K, and surface gravities of log(g (cm/s$^{2}$)) $=$ 4.23--4.37. The second set of models was obtained from the CMFGEN website\footnote{\url{https://sites.pitt.edu/~hillier/web/CMFGEN.htm}, \\'A Grid of O star CMFGEN Models', accessed 2023/11/02.} in order to include higher mass O-type stars in our method. We select 4 models spanning luminosities of log(L/L$_{\odot}$) = 4.95--6.1 and temperatures of T$_{\mathrm{eff}}$ = 35000--40000K. We then estimate their masses by comparing the luminosities and temperatures to MIST stellar atmosphere models \citep{Dotter.A.2016.MISTPaperI,Choi.J.2016.MISTPaperII}, and calculate masses of 20--60 M$_{\odot}$. 

We combine these models by scaling the flux of all models to the distances of the SMC or LMC (62 or 50 kpc, respectively, \citealt{Graczyk.D.2014.SMCDistance,Pietrzynski.G.2013.LMCDistance}), and combining each OB-type MS model flux with each YSG model flux. 

To facilitate comparison with observations and to understand how different levels of extinction can impact candidate selection, we then use the python package \texttt{dust\_extinction}\footnote{\url{https://dust-extinction.readthedocs.io/en/stable/}} \citep{Gordon.K.2023.dustextinction} to apply reddening to these combined models. We use the \citet{Gordon.Karl.2003.MCExtinction} extinction curves for the Magellanic Clouds with R$_{\mathrm{V}}$ = 2.74/3.41 for the SMC/LMC, respectively (using the LMC `average' curve), and consider the full range of A$_{\mathrm{V}}$ values for cool stars from \citet{Zaritsky.D.2002.MCPSSMC,Zaritsky.D.2004.MCPSLMC}. As a `typical' extinction value for the plots in this paper, we adopt A$_{\mathrm{V}}$ values of 0.22/0.38 for the SMC/LMC, respectively. These values are from \citet{DroutGotberg.2023.IntStrippedStars}, who found those values to be the best for correctly aligning the theoretical ZAMS with UV colors of observed stars in the Clouds. These values are close to those measured by \citet[0.19/0.44]{Zaritsky.D.2002.MCPSSMC,Zaritsky.D.2004.MCPSLMC}, and \citet[0.25/0.44]{Massey.P.1995.MagellanicCloudsExt}.

\begin{figure*}
    \centering
    \includegraphics[width=0.95\textwidth]{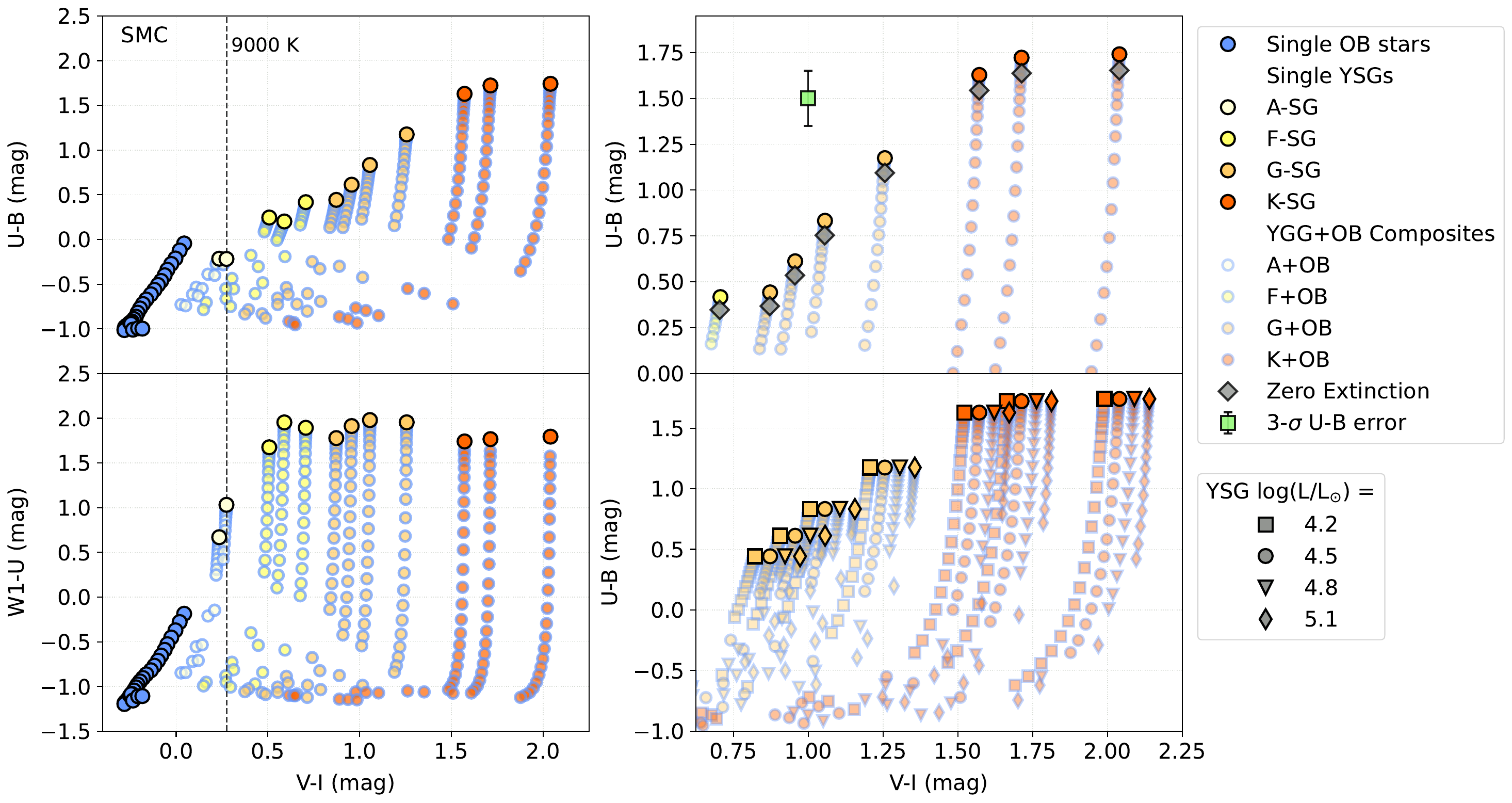}
    \caption{Color-color diagrams displaying the stellar atmosphere models used to design photometric criteria for identifying YSG+OB binaries. Pickles YSG models are shown as large off-white, yellow, light orange, and dark orange (A-, F-, G-, and K-type, respectively) circles, outlined in black. Single OB-type MS stars are shown as blue circles in the lower left. The combined models are shown as smaller circles with pale blue outlines. SMC models are shown here, LMC models show the same behaviour. \textit{Left panels:} Colors shown are V$-$I vs U$-$B (top) and W1$-$U (bottom), with a YSG luminosity of log(L/L$_{\odot}$) = 4.5. The dashed grey vertical lines correspond to T$_{\mathrm{eff}}$=9000 K. \textit{Top right:} For U$-$B colors, the effect of zero extinction is shown with a grey diamond, and the 3-$\sigma$ uncertainty in U$-$B (assuming an uncertainty of 0.05 mag) is shown with a green square and error bar. \textit{Bottom right:} The effects of 4 different YSG luminosities are shown: log(L/L$_{\odot}$) = 4.2 (square), 4.5 (circle), 4.8 (triangle), 5.1 (diamond). Luminosities other than 4.5 are horizontally offset for clarity.}
    \label{fig:combinedmodels}
\end{figure*}

In Figure~\ref{fig:fracflux} we show the relative contributions of the binary components to the flux in different photometric bands for our entire model grid. Each small box in this plot represents one combined model for a given YSG luminosity (log(L/L$_{\odot}$) = 4.0, 4.5, and 5.0, for the top, middle and bottom rows of plots, respectively), YSG spectral type (horizontal axis), and OB-star mass (vertical axis). Each box is colored to indicate the percentage of flux that the OB-type star contributes to the total combined flux in the synthetic photometric band indicated in each column (W1, U, B, V, and J, left to right). These photometric bands were chosen to span the wide range of wavelengths that we use in \S\ref{sec:criteria}.

From this, we see that the OB-type star contributes effectively zero flux for all models in the J-band except for the most extreme pairs (a low luminosity A0 star with a 50M$_{\odot}$ O star companion). Even in the V-band, the companion star contributes minimally in a vast majority of models, with contributions of $\gtrapprox$30\% appearing only for the highest mass OB-star models. The OB-star contributes increasingly more flux at lower YSG luminosities, later YSG spectral types (cooler temperatures), higher OB-star masses, and bluer bands. For UV filters, the OB-star flux dominates nearly 100\% of the flux in half or more of the parameter space. However, even for low luminosity YSGs and UV filters, there are certain combinations of YSG temperature and OB-star mass where the YSG still contributes nearly all of the flux.

\subsection{Identifying YSG Binary Candidates in Color-Color Space}\label{sec:criteria}

Here we describe the details of our method for photometrically identifying YSG+OB binary candidates. Broadly, we rely on the models constructed in \S\ref{sec:combinedmodels}, to identify stars that have a blue or UV color excess relative to the colors of single YSG models. To aid in this process, in Figure~\ref{fig:combinedmodels} we show the 324 final combined, extincted models---along with the original YSG and OB-MS models---in multiple color-color diagrams. We show both an example that relies on only optical colors (U$-$B vs V$-$I; top left) and one that incorporates a UV magnitude (W1$-$U vs V$-$I; bottom left). In these panels, the YSG models have log(L/L$_{\odot}$) = 4.5. SMC models are plotted, but the LMC models show the same overall behavior, with a slight shift in V$-$I colors due to metallicity and dust differences.

\subsubsection{Optical Color-Color Diagrams}\label{sec:opticalccc}

Although the combined models show that OB-type companions in YSG binaries are more clearly visible with UV photometry, optical photometry is much more readily available than UV in the Clouds. We therefore first investigate the effect of an OB-star companion on YSGs in optical color space. 

In the optical color-color diagram (Figure~\ref{fig:combinedmodels}, top left panel), we can clearly see that the YSG+OB binary systems exist in a unique region of color-color space. For a given YSG, adding a progressively larger OB companion will cause the system to appear at bluer and bluer U-B colors. Notably, for all but systems with the most massive O-star companions, this process happens at nearly constant V-I color. This is indicative of the fact that moderate mass blue companions do not significantly contribute to the flux in red optical bands.

As a result, there is a clear separation between single YSGs and many YSG+OB binaries in this phase space. However, we note that for some companion masses, the predicted photometry of the combined models lies within a few tenths of a magnitude of the nearest single YSG model. This is particularly true for higher temperature YSGs, where many models can be found at similar U$-$B colors. Given that we will identify candidate YSG+OB binaries as systems that are not ``too close'' to the nearest single YSG model in this phase space, we consider the impact of three different effects on the minimum OB companion mass required to produce a \emph{detectable} blue excess relative to a single YSG as a function of temperature: i) photometric uncertainty, ii) dust extinction, and iii) the YSG luminosity.
\begin{enumerate}
\setlength\itemsep{-0.1em}
    \item A high photometric uncertainty could lead to a YSG+OB binary, particularly those with small B-star masses, to be confused with a single YSG, and vice-versa. In the top right panel of Figure~\ref{fig:combinedmodels} we show an error bar that represents three times the median uncertainty for U$-$B measurements in MCPS ($\langle \sigma_{\rm{U-B}} \rangle = 0.05$ mag). On average, detecting a color excess at a 3$\sigma$ level would require a shift relative to the single YSG models of this amount.
    \item We assume one extinction value for each Magellanic Cloud, but individual stars will have a range of dust extinction values. Dust extinction more heavily attenuates blue flux, so a U$-$B color of a YSG will be bluer at low extinction values (and thus more easily confused with a YSG+OB system with moderate extinction). We take zero extinction as the case where a single YSG will be the bluest in U$-$B. We demonstrate the shift that this would produce relative to our baseline models in the top right panel of Figure~\ref{fig:combinedmodels}.
    \item In binaries with higher luminosity YSGs, an OB companion of a given mass will contribute a smaller fraction of the overall flux, and thus a smaller relative amount of excess blue color. In the bottom right panel of Figure~\ref{fig:combinedmodels}, we demonstrate the effect of the YSG luminosity on the location of YSG+OB binaries in this phase space (models are offset horizontally for clarity). The more luminous the YSG, the more compact the range of U$-$B values for YSG+OB binaries.
\end{enumerate}

\begin{figure}
    \centering
    \includegraphics[width=0.45\textwidth]{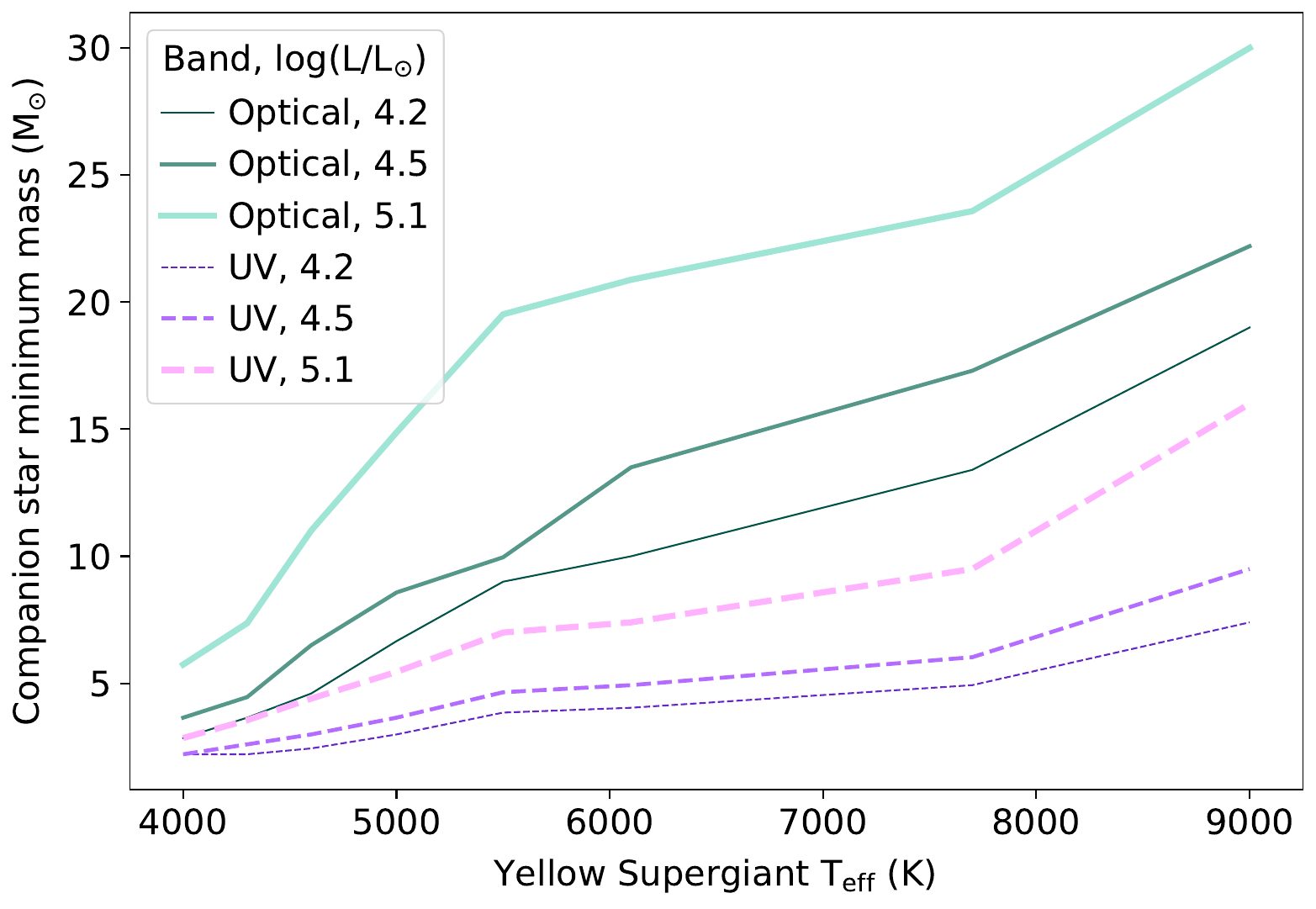}
    \caption{YSG T$_{\mathrm{eff}}$ vs minimum OB star mass. Each line shows, as a function of YSG temperature, the minimum mass of OB companion star required for the binary to be discernible from a single YSG at the same temperature, specifically requiring the binary model to be 0.15 mag bluer in U$-$B or W1$-$U. Optical colors are solid green lines, UV colors are dashed violet lines. YSG luminosities of log(L/L$_{\odot}$)= 4.2, 4.5, and 5.1 are shown as lines of increasing thickness.}
    \label{fig:compmassteff}
\end{figure}

Comparing the first two cases, we find that for typical U$-$B MCPS uncertainties ($\langle \sigma_{\rm{U-B}} \rangle = 0.05$ mag) requiring a 3$\sigma$ color excess would encompass the region of the zero extinction models and is the dominant of these two effects. In Figure \ref{fig:compmassteff} we plot the minimum companion mass of a YSG+OB binary that would be discernible from a single YSG at the same temperature, if we require the combined binary model to have a U$-$B color that is at least 0.15 mag bluer than that of the single YSG (thus representing a 3$\sigma$ excess for typical MCPS uncertainties). For this plot we assume `typical' extinction values described in \S\ref{sec:combinedmodels} and a range of YSG luminosities (log(L/L$_{\odot}$) = 4.2, 4.5, 5.1). When using only optical colors (solid green lines), we see the minimum OB companion mass ranges from 3-6M$_{\odot}$ for cool YSGs ($\sim$4000 K) to 19-30M$_{\odot}$ for hotter YSGs ($\sim$9000 K), depending on the YSG luminosity. Thus, the mass ratios of YSG+OB binaries this method is sensitive to depends strongly on both the YSG luminosity and temperature.

\subsubsection{Ultraviolet-Optical Color-Color Diagrams}\label{sec:uvccc}

In the bottom left panel of Figure~\ref{fig:combinedmodels}, we plot V$-$I vs. W1$-$U. When UV magnitudes are considered, the majority of combined models are even more greatly separated from their associated single YSG models. This is expected given OB stars are relatively more bright in UV than YSGs. In Figure \ref{fig:compmassteff}, the minimum companion mass of discernible YSG+OB binaries when using UV colors (dashed violet lines) are significantly lower than in optical. To make this fiducial plot we require a that a given combined model have a W1$-$U color that is 0.15 mag bluer than the closest single YSG model. This will be compared to typical photometric errors that can be acheived with \emph{Swift} photometry in Section~\ref{sec:uvcol-results}, below. The minimum companion mass that is descernable when utilizing UV photometry ranges from 2-3 M$_{\odot}$ at 4000 K to 7-16 M$_{\odot}$ at 9000 K, depending on YSG luminosity. UV photometry therefore allows us to identify YSG+OB systems with much smaller companion masses than optical photometry alone.

\section{Identifying Yellow Supergiant Binaries}\label{sec:results}

Here we present the results of using the unique regions of color-color space identified in \S\ref{sec:method} to identify YSG+OB binary candidates in the Magellanic Clouds.

\subsection{Identifying Yellow Supergiants}\label{sec:idYSG}

To achieve our goal of identifying YSGs with binary companions, we must first identify YSG candidates in the Magellanic Clouds. Here we describe our selection criteria and foreground contamination removal. We generally follow the methods of \citet{Drout.M.2009.YSGinM31,Drout.Maria.2012.RSG.YSG.M33} and \citet{Neugent.K.2010.YSGinSMC,Neugent.K.2012.RSG.YSG.LMC}. 

To encompass the extent of the SMC, we first select all sources from 2MASS \citep{2MASS.Skrutskie.2006} within 2.25$^{\circ}$ of 0h55m11s $-$72$^{\circ}$57$\farcm$00$\farcs$ (J2000), which is located in the central region of the SMC. We find 252,472 sources. For the LMC, we select all sources within 4.5$^{\circ}$ of 5h15m20s $-$69$^{\circ}$20$\farcm$10$\farcs$ (J2000), giving us 1,291,696 sources. We apply our typical extinction corrections from \citet{DroutGotberg.2023.IntStrippedStars} of A$_{\mathrm{V}}$ = 0.22 (SMC) or 0.38 (LMC). This corresponds to E(B$-$V) = 0.08 (SMC, R$_{\mathrm{V}}$=2.74) or 0.11 (LMC, R$_{\mathrm{V}}$=3.43).

\subsubsection{Photometric Identification of YSGs}\label{init_cc}

\begin{figure}
    \centering
    \includegraphics[width=0.45\textwidth]{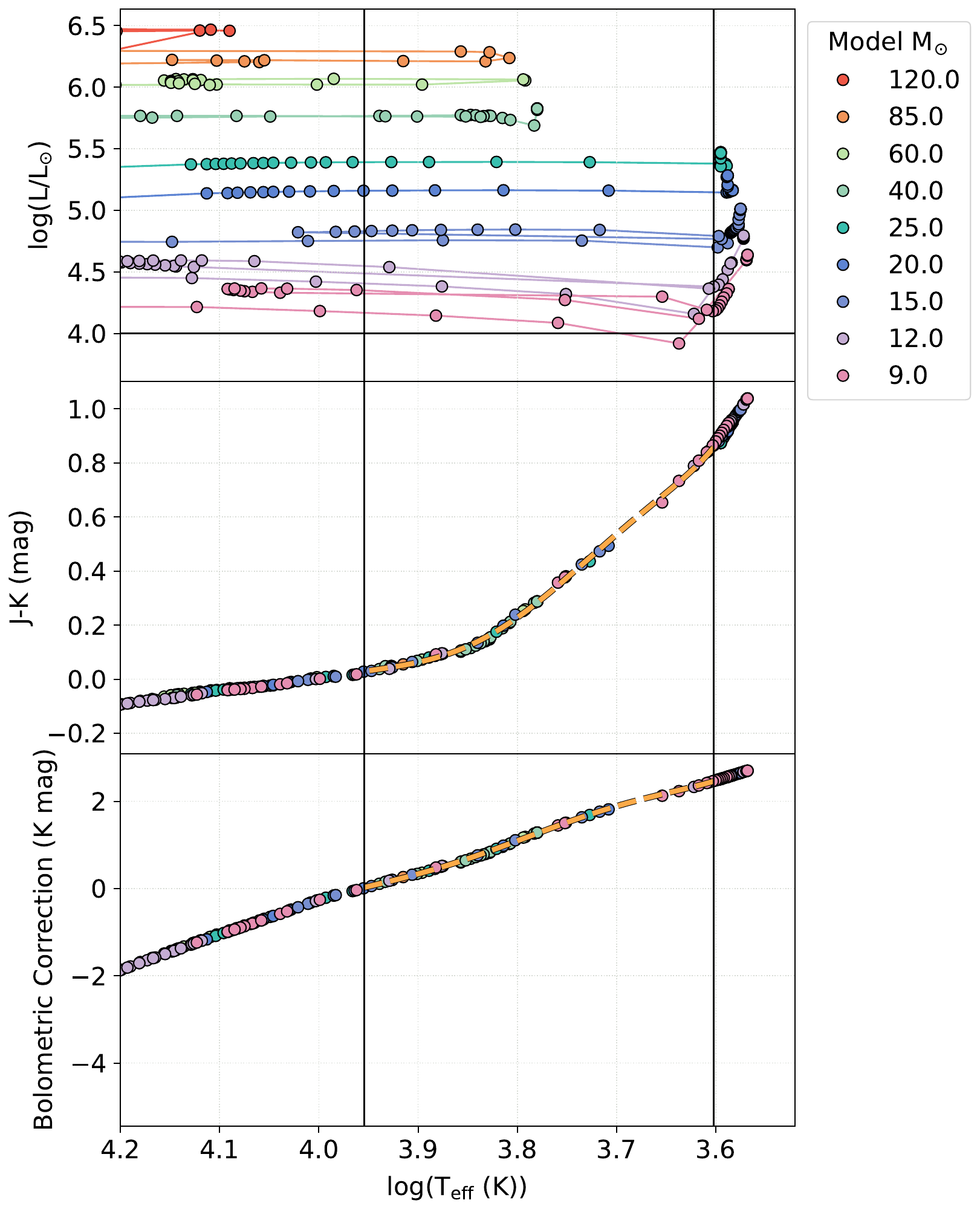}
    \caption{\textit{Top:} Hertzsprung-Russell diagram showing the Geneva stellar evolutionary models \citep{Lejeune.T.2001.GenevaPhotModels} at metallicity Z=0.008 for the LMC. A range of colors denotes mass. Black solid lines denote the region of interest (log(L/L$_{\odot}$) $\geq$ 4.0 and 4000 $\leq$ T$_{\mathrm{eff}}$ $\leq$ 9000 K). \textit{Middle:} Effective temperature vs J$-$K color for the same models. Solid vertical lines denote the same temperature range as above, and a dashed orange line shows the fitted relationship. \textit{Bottom:} Same as the middle panel, but for effective temperature vs K-band bolometric correction.}
    \label{fig:ysgidinfo}
\end{figure}

To select stars within the temperature and luminosity range of YSGs, we use the J$-$K and K-band magnitudes from 2MASS. We use infrared colors because, as demonstrated in Figure~\ref{fig:fracflux}, infrared magnitudes will be unaffected by the presence of all but the most massive (and therefore most rare) OB-type star companions. In the infrared, nearly all single and binary YSGs will appear the same. 

To determine which infrared colors and magnitudes correspond to the temperature and luminosity cuts of log(L/L$_{\odot}$) $\geq$ 4.0 and 4000 $\leq$ T$_{\mathrm{eff}}$ $\leq$ 9000 K (our adopted definition of a YSG; see \S\ref{sec:ysgdef}), we use the photometric information distributed by \citet{Lejeune.T.2001.GenevaPhotModels}, who use the synthetic stellar spectra libraries of \citet{Lejeune.T.1997.SynthSpecLib} and \citet{Westera.P.1999.SynthSpecLibZ} to calculate synthetic photometry for the Geneva stellar evolutionary models. First, we select models with Z = 0.004 (SMC) or Z = 0.008 (LMC) and M $\geq$ 9M$_{\odot}$ (since the 9M$_{\odot}$ model traverses the Hertzsprung-Russell diagram at the limit of log(L/L$_{\odot}$) $\approx$ 4.0). We then apply the distance correction for the SMC/LMC, and transform the generic J and K Geneva magnitudes to 2MASS J and K$_{\mathrm{s}}$ magnitudes \citep{Carpenter.J.2001.ColorTransformations2MASS}. 

Next, we define a relationship (a 6-order polynomial, following \citealt{Neugent.K.2010.YSGinSMC,Neugent.K.2012.RSG.YSG.LMC}) between J$-$K and T$_{\mathrm{eff}}$ to select stars with 4000 $\leq$ T$_{\mathrm{eff}}$ $\leq$ 9000 K. Then, using the K$_{\mathrm{s}}$ values and luminosities in the Geneva models, we again fit a polynomial to determine the bolometric correction (BC) as a function of temperature, where BC = $\mathrm{M_{bol}} -$ K$_{\mathrm{s}}$ and $\mathrm{M_{bol}} = -2.5\times\mathrm{log(L/L_{\odot})}+4.75$. We then use these BCs to determine the K-band magnitude that corresponds to the minimum luminosity of log(L/L$_{\odot}$) = 4.0 as a function of J$-$K (i.e., as a function of temperature). In Figure~\ref{fig:ysgidinfo} we show the LMC Geneva models and the two fitted relationships.

We now comment on the robustness of temperature estimates made using J$-$K color over our selected temperature range. As shown in the middle panel of Figure~\ref{fig:ysgidinfo}, J$-$K is very sensitive to temperatures between 4000 and 7000 K. Within this range, photometric uncertainties of $\sim$0.1 mag in J$-$K would result in temperature uncertainties of less than 100 K. However for temperatures between 7000 and 9000 K, the relationship flattens as the peak of the spectral energy distribution shifts to shorter wavelengths and J$-$K becomes much less sensitive to T$_{\mathrm{eff}}$. In this range, J$-$K uncertainties of 0.15 mag result in temperature uncertainties of 1000K at the low end to 6000 K at the extreme high end (i.e. a 15,000K star may be incorrectly viewed as a 9000K star). We comment on the impact of this uncertainty on our sample of hot YSG candidates in \S\ref{sec:contaminants}.

After using these color/temperature and magnitude/luminosity relationships to select candidate YSGs with log(L/L$_{\odot}$)$\geq$4.0 and 4000$\leq$T$_{\mathrm{eff}}$$\leq$ 9000 K, we are left with 3,236 stars in the SMC and 10,129 in the LMC.

\subsubsection{Removal of Foreground Sources}\label{sec:gaia}

The area of color-magnitude space that YSGs within Local Group galaxies occupy is also heavily contaminated by foreground dwarfs \citep{Drout.M.2009.YSGinM31,Drout.Maria.2012.RSG.YSG.M33,Neugent.K.2010.YSGinSMC,Neugent.K.2012.RSG.YSG.LMC}. We follow a method based on that of \citet{Gaia2018} and described in detail in \citet{O'Grady.A.2020.superAGBidentification,OGrady.A.2023.sAGBTZOAnalysis}, to minimize foreground dwarf contamination using \emph{Gaia} astrometry. After cross-matching, 180/479 stars in the SMC/LMC did not have matches in the Gaia DR3 database \citep{Gaia.2023.DR3}, and we do not consider them further. We then remove all sources with parallax over parallax error $\pi/\sigma_{\pi}>$ 5, and compare the proper motions of the remaining sources to a covariance matrix describing the distribution of proper motions of $\sim$ 1,000,000 highly probable SMC/LMC members. Stars that fall outside the region of proper motion space containing 99.5\% of these probable Cloud members, corresponding to $\chi^{2} >$ 10.6, are removed as likely foreground stars. This astrometric cut leaves 541/729 sources in the SMC/LMC, implying a foreground contamination level of 82-92\%, consistent with previous works \citep{Drout.M.2009.YSGinM31,Drout.Maria.2012.RSG.YSG.M33,Neugent.K.2010.YSGinSMC,Neugent.K.2012.RSG.YSG.LMC}.

We also investigate how many foreground yellow dwarfs with proper motions similar to those of the Clouds may still be present in our sample after applying this method. We apply the same color and kinematic cuts to stars within a control field with a radius of 10$^{\circ}$ at a similar Galactic latitude as the Clouds, centered on ($\alpha, \delta$) (J2000) = (22:13:28.17, $-$63:06:55.92). From a field containing over 750,000 stars, we find less than 0.5\% of stars could be mistaken for YSGs with proper motions consistent with either SMC or LMC membership. Thus, we anticipate only 2-3 candidate YSGs in either Cloud to be foreground dwarf contaminants.

\subsubsection{Photometry Cross Matching}\label{sec:crossmatch}

In order to use the photometric selection methods described in \S\ref{sec:criteria}, we require optical and (if available) UV photometry for our sources. For optical photometry, we cross-match the YSG sample to the Magellanic Cloud Photometric Survey \citep[MCPS,][]{Zaritsky.D.2002.MCPSSMC,Zaritsky.D.2004.MCPSLMC}, which provides U, B, V and I band magnitudes for several million sources in the SMC/LMC. This yields 475/591 matches in the SMC/LMC. Of the 66/138 that did not match, 13/1 fall outside of the MCPS footprint, while 53/137 are within the MCPS footprint but do not match. Additionally, 54/73 stars ($\sim$11-12\%) are missing either U- or I-band data, which is essential for the remainder of this work. Thus the final count of YSG candidates with complete MCPS data is 421/518 in the SMC/LMC.

For the UV, we use data from the Swift Ultraviolet Survey of the Magellanic Clouds \citep[SuMAC,][]{Hagen.L.2017.SwiftSUMAC}. SUMaC images were taken in the W2, M2, and W1 ultraviolet bands, covering $\sim$9 and 3 deg$^{2}$ in the LMC and SMC, respectively. Photometry was computed for stars in the SUMaC images by \citet{Ludwig23} by performing 
forced PSF photometry at the location of known stars from the MCPS catalogs using the forward modeling code \textit{The Tractor} \citep{Lang.D.2016.Tractor,Lang.D.2016.WiseTractor}. This technique was used in order to mitigate crowding effects; further details of this forward modeling forced photometry are given by \citet{Ludwig23} and \citet{DroutGotberg.2023.IntStrippedStars}. Cross matching our candidates with complete optical data to the final catalog of \citet{Ludwig23} yields 277/167 matches in the SMC/LMC, but 5/14 stars did not have detections in the W1 filter, which is necessary for our method, leaving a total of 272/153 stars with the needed UV data. Of the 144/351 that did not match, 
99/323 fall outside of the SuMAC footprint in the Clouds, while 45/28 are within the SuMAC footprint but do not match.

For the stars within the SuMAC footprint but without matches in the final \citet{Ludwig23} catalog, we examine the SuMAC images to determine why. We find that: (i) 27 were located in regions of the images that were masked during the analysis process by \citet{Ludwig23}, (ii) 26 were too faint to yield a $>$3$\sigma$ detection in the W1 band, and (iii) 20 were excluded from the final catalog of \citet{Ludwig23} because there were high residuals left in a 5$\arcsec$ aperture surrounding the source after subtracting the best-fit model for all sources.

For the first set of objects, masking was performed in regions that were either highly clustered or featured a bright source, such that prominent coincidence-loss artifacts appeared on the images. Robust magnitudes cannot be recovered from the Swift-UVOT images for these sources. However, we note that for approximately half of these sources, the YSG itself appeared to be very bright in the UV; which is not expected for isolated YSGs. For the second set of objects, we use a combination of the outputs from \emph{Tractor} and aperture photometry using the \texttt{uvotsource} tool distributed by \textsc{heasarc}\footnote{\url{https://swift.gsfc.nasa.gov/analysis/UVOT_swguide_v2_2.pdf}} to estimate 3$\sigma$ upper limits for the YSGs when possible. Finally, we visually inspect all of the sources in the third group. In some cases, the high residuals were an indication the the source was in a highly clustered region or that the adopted PSF was not a good match to that image. However, for 19 objects we found that the residuals were only slightly above the threshold applied by \citet{Ludwig23}, and were likely due to a brighter source located $>$3\arcsec\ away from the target of interest. In these cases we deem the UV magnitudes from \emph{Tractor} reliable enough for our purposes and include them in the following analysis.

In summary, we supplement the catalog of \cite{Ludwig23} with UV magnitudes for 15/4 sources in the SMC/LMC, as well as 7/5 for which we have obtained upper limits on the true flux from the system. After adding these sources, we are left with 272/156 sources with sufficient data to apply our UV-optical method for identifying YSG binaries. 

\subsubsection{Cleaning Spectral Energy Distributions}\label{sec:sed-clean}

Since YSG+OB binary candidates will be identified based on their optical and UV colors, we carefully examine the spectral energy distributions (SEDs) of YSG candidates to remove any with bad photometry. Using the YSG and YSG+OB combined models from \S\ref{sec:combinedmodels} as guides, we remove stars where any MCPS magnitude (or, if available, UV magnitude from \citealt{Ludwig23}) implies a non-physical SED. For most bands, we exclude sources where a given band is more than 1.5 AB magnitudes brighter or fainter than an adjacent band. The exception is U-band, where the cooler YSG models show variation of up to 2.5 magnitudes relative to B-band. This process removes 39/37 stars in the SMC/LMC, respectively, leaving us with 382/481 with complete data and clean SEDs. For stars with UV data, an additional 1 star is removed in both the SMC and LMC where the W1-band was more than 2.5 magnitudes brighter or fainter than the U-band.

\subsubsection{YSG Candidate Sample Summary}

\begin{deluxetable}{l|rr}
\centering
\tablecaption{YSG Sample Construction\label{ysg_selection}}
\tablehead{\multicolumn{1}{c|}{Criterion} & \colhead{SMC} & \colhead{LMC}}
\startdata
2MASS Sources & 252,472 & 1,291,696 \\ 
J$-$K and K Cuts & 3,236 & 10,129 \\
Gaia Cross-Match & 3,056 & 9,650 \\ 
Gaia Astrometry Cut & 541 & 729 \\ 
\hline
\multicolumn{3}{c}{Optical Data}\\ 
\hline
MCPS Cross Match & 475 & 591 \\
With complete data & 421 & 518 \\
Cleaned SEDs & 382 & 481 \\
\hline
\multicolumn{3}{c}{UV Data}\\
\hline
SuMAC Cross Match & 277 & 167 \\
With UV W1 data & 272 & 153 \\
With extra sources & 286 & 157 \\
Cleaned SEDs & 272 & 156 
\enddata
\end{deluxetable}

\begin{figure*}
    \centering
    \includegraphics[width=0.9\textwidth]{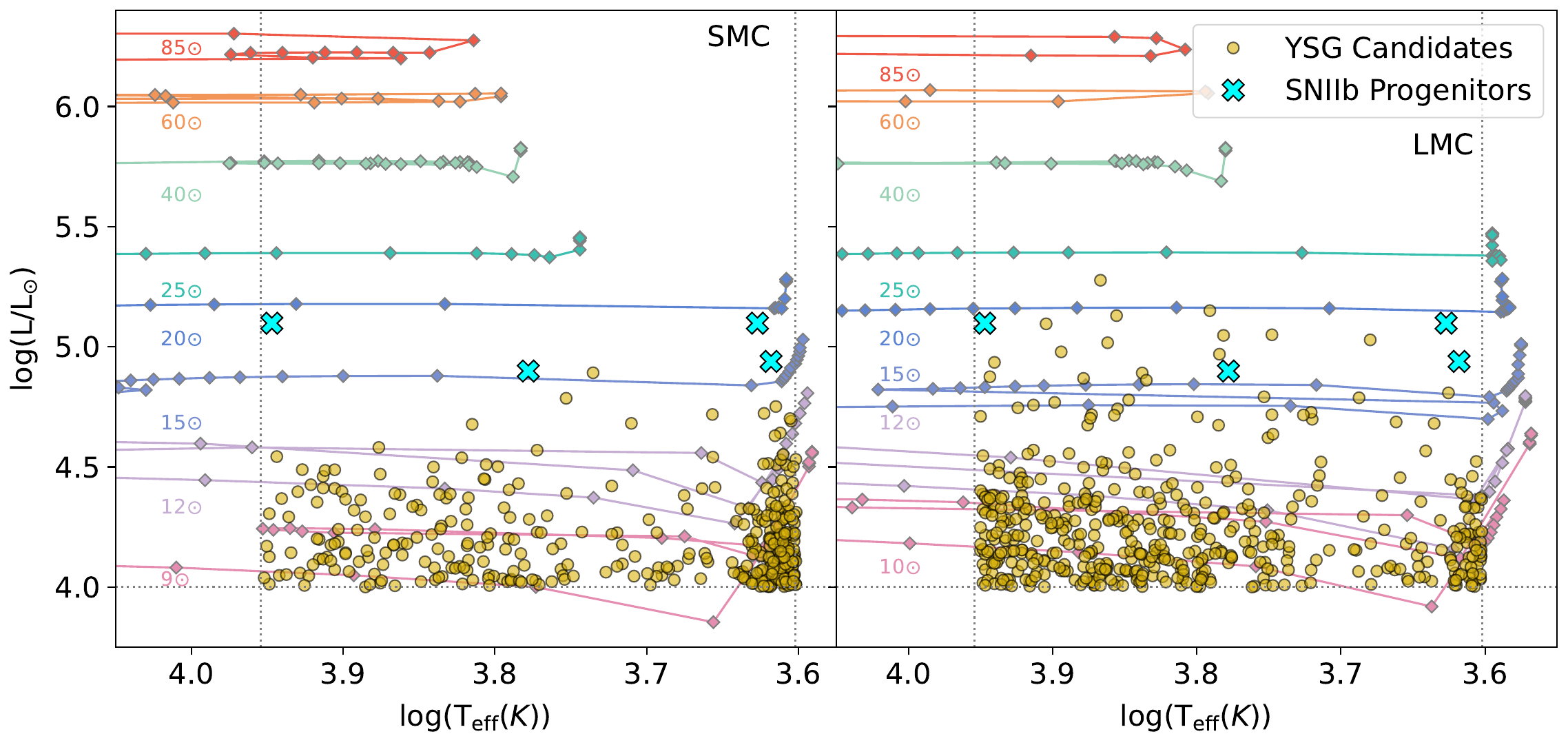}
    \caption{\textit{Top:} Hertzsprung-Russell diagrams for the SMC (left) and LMC (right) showing the final YSG candidate sample as gold circles. The limits of log(L/L$_{\odot}$)$\geq$4.0 and 4000$\leq$T$_{\mathrm{eff}}\leq$9000 K are shown as grey dotted lines. As well, the Geneva stellar evolutionary models \citep{Lejeune.T.2001.GenevaPhotModels} at metallicities Z=0.004/0.008 for the SMC/LMC are plotted, with masses indicated. Finally, examples of partially-stripped-envelope progenitors from \citet{Eldridge.J.2013.IIbYSGPlot} are shown as light blue crosses.}
    \label{fig:hrd_ysgcand}
\end{figure*}

In Table \ref{ysg_selection}, we summarize the progression of cuts made to arrive at our final sample of YSG candidates in the Magellanic Clouds: 541 in the SMC and 729 in the LMC, with 382/481 in each galaxy with all MCPS magnitudes available and clean SEDs. The final sample of YSG candidates is shown on a Hertzsprung-Russell diagram in Figure~\ref{fig:hrd_ysgcand}; temperatures and luminosities were determined by comparing J- and K$_{\mathrm{s}}$ magnitudes to the Geneva models by the process described in \S\ref{init_cc}. For context, we also show the location of several observed YSG progenitors of partially stripped-envelope supernovae from Figure 12 of \citet{Eldridge.J.2013.IIbYSGPlot}.

We compare our final population of YSG candidates to that of \citet{Neugent.K.2010.YSGinSMC,Neugent.K.2012.RSG.YSG.LMC}. They searched for YSGs with M$\geq$12M$_{\odot}$ with a similar method and found 192/317 probable YSGs in the SMC/LMC, respectively. Our YSG candidate population is larger mainly due to the inclusion of fainter stars, consistent with expectations for 9--11M$_{\odot}$ YSGs based on evolutionary models. Small differences also originate from how we handled foreground dwarf removal.

Of the 192/317 YSG candidates in the \citet{Neugent.K.2010.YSGinSMC,Neugent.K.2012.RSG.YSG.LMC} population, when compared to our 541/729 candidates before cross-matching to MCPS and making SED cuts, 68/119 are missing from our final population in the SMC/LMC, respectively. These missing \citet{Neugent.K.2010.YSGinSMC,Neugent.K.2012.RSG.YSG.LMC} YSGs were removed from our final sample for the following reasons: 
\vspace{-0.05in}
\begin{enumerate}
\setlength\itemsep{-0.1em}
    \item 3 SMC YSGs are located in the wing of the SMC, which we do not consider.
    \item 49/108 YSGs were removed through our photometric cuts on J$-$K and K, which were designed to select stars in our temperature/luminosity region of interest (see \S\ref{init_cc}).
    \item 16/11 more of the \citet{Neugent.K.2010.YSGinSMC,Neugent.K.2012.RSG.YSG.LMC} YSGs were then removed from our sample when removing likely foreground dwarfs based on their \emph{Gaia} astrometry (see \S\ref{sec:gaia}).
\end{enumerate}
\vspace{-0.05in}

For the stars removed through photometric cuts, these missing YSGs stem from a) slight differences in how we and \citet{Neugent.K.2010.YSGinSMC,Neugent.K.2012.RSG.YSG.LMC} handle extinction, and b) \citet{Neugent.K.2010.YSGinSMC,Neugent.K.2012.RSG.YSG.LMC} using more lenient color cuts in J$-$K. Using the J$-$K values in the catalogs provided by \citet{Neugent.K.2010.YSGinSMC,Neugent.K.2012.RSG.YSG.LMC} and our J$-$K-T$_{\mathrm{eff}}$ relation, the \citet{Neugent.K.2010.YSGinSMC,Neugent.K.2012.RSG.YSG.LMC} YSGs range in temperature from 3900 - 10300 K, wider than our range of 4000 - 9000 K. If we apply our extinction correction and color cuts to the missing stars, the only ones which pass the cuts are the 3 SMC wing stars and the 16/11 that are later cut by our Gaia analysis.

We also compare our YSG candidate sample to the fast yellow pulsating supergiants (FYPS), a potential new class of luminous pulsating stars identified by \citet{DornWallenstein.T.2020.FYPS} (though see also \citealt{Pedersen.M.2023.FYPSTessContam}). Of the 5 FYPS identified by \citet{DornWallenstein.T.2020.FYPS}, 1 is within our YSG candidate sample (2MASS J04505585-6925525). The 4 others are not in our final sample because they are not in MCPS.

\subsection{Binary Candidate Selection}\label{sec:id_binary_cand}

We now search for evidence of excess blue and/or UV flux that may indicate the presence of binary companions for some of the YSGs identified in \S\ref{sec:idYSG}. We use both optical-only and optical-UV color-color diagnostics described in \S\ref{sec:criteria} (the latter of which is available for only a subset of the YSG sample). 

In the sections below, we will broadly define a star as having a blue or UV color excess (and thus being a YSG+OB candidate) when it's U$-$B or W1$-$U color is more than three times the typical photometric uncertainty \emph{bluer} in U$-$B or W1$-$U, for optical-only and optical-UV diagnostics respectively, than a single YSG model at the same V$-$I color. This requirement is chosen to ensure that the excess (relative to the Pickles models) is measured at at least 3$\sigma$. Such stars will appear below the single YSG models in Figure \ref{fig:combinedmodels}. To make this assessment for stars at a range of V$-$I color, gaps between the YSG models are linearly interpolated.

\begin{figure*}
    \centering
    \includegraphics[width=0.9\textwidth]{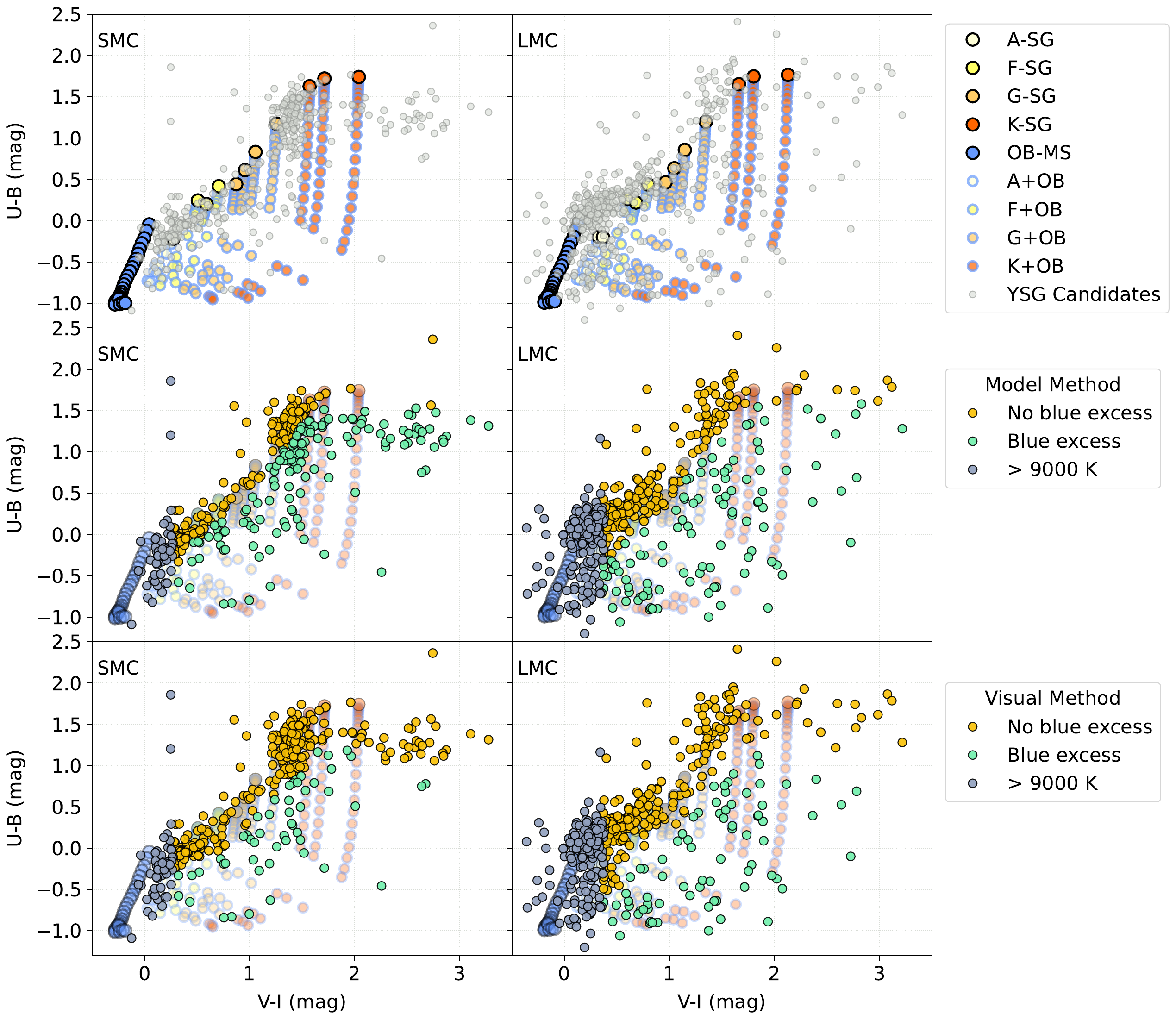}
    \caption{Color-color diagrams displaying observed candidate YSGs for V$-$I vs U$-$B, at SMC (left) and LMC (right) metallicities. The stellar atmosphere models from Figure~\ref{fig:combinedmodels} are also shown. \textit{Top:} Shown in grey are YSG candidates, selected as described in \S\ref{sec:idYSG}. \textit{Middle:} The YSG candidates are classified as described in \S\ref{sec:optcol-model}. Gold circles are YSGs with no photometric evidence of a blue excess. Green circles are YSGs with photometric evidence of a blue excess, and are possibly binary candidates. Dark blue-grey circles are YSGs that have V$-$I colors corresponding to T$_{\mathrm{eff}}>$9000 K. \textit{Bottom:} YSG candidates are classified by a `visual over-density' method, described in \S\ref{sec:optcol-visual}. Colors are the same as the panel above.}
    \label{fig:opticalcand}
\end{figure*}

\begin{figure*}
    \centering
    \includegraphics[width=0.9\textwidth]{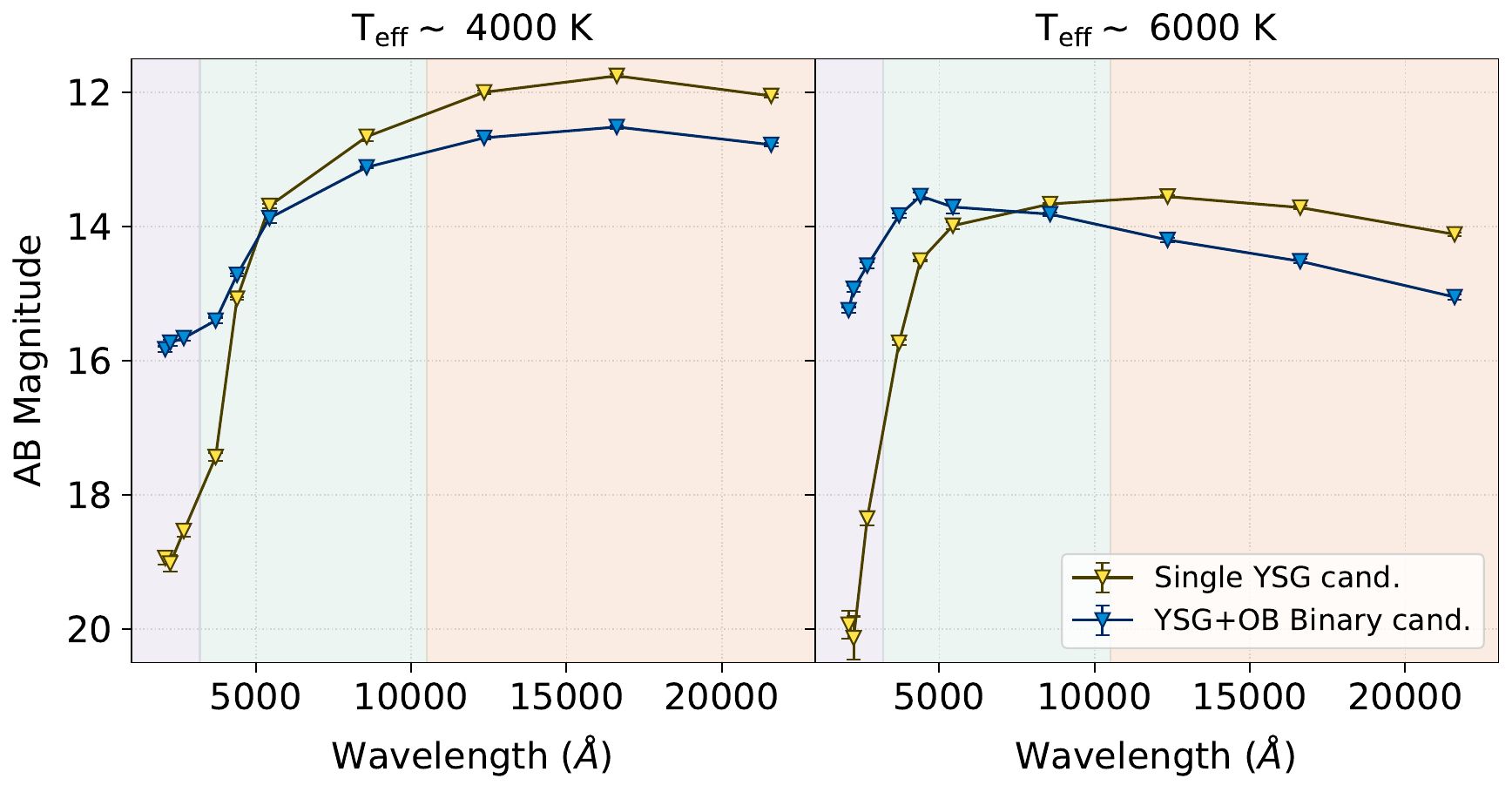}
    \caption{Spectral energy distributions of four YSG candidates in our sample. Four stars are shown: two have T$_{\mathrm{eff}}\sim$ 4000 K (left panel) and two have T$_{\mathrm{eff}}\sim$ 6000 K (right panel), and two are likely single YSGs (yellow) while two are YSG+OB binary candidates (blue). Wavelength ranges corresponding to the SUMaC, MCPS, and 2MASS surveys are shaded violet, green, and orange, respectively.}
    \label{fig:sed_example}
\end{figure*}

\subsubsection{Optical Colors - Model Method}\label{sec:optcol-model}

We first use the optical-only color-color diagnostic plot presented in \S\ref{sec:criteria} to separate the observed sample into three broad classes:

\begin{enumerate}
\setlength\itemsep{-0.1em}
    \item \textit{YSGs too hot for definitive statement:} We first categorize any star in our YSG sample with a V$-$I color bluer than the Pickles A2I model (left of the dashed line in Figure \ref{fig:combinedmodels}) as a 'too hot YSG'. The Pickles A2I model has a temperature of $\sim$9000 K, which was the upper limit we adopted for our definition of YSGs (\S~\ref{sec:ysgdef}). While we nominally only selected YSGs with temperatures below this, some stars clearly appear in this region of parameter space. This could occur due to i) differences between the \citet{Lejeune.T.1997.SynthSpecLib} stellar atmosphere models (used to select YSGs) and Pickles models (which we now compare to), and/or ii) differences in the sensitivity of J$-$K vs V$-$I to temperature at the hot end our temperature range (discussed in \S~\ref{init_cc}).
    \item \textit{No observable blue/U$-$B excess:} Stars that are close to the single YSG models and do not show signs of excess blue flux in optical colors. The median U$-$B error on our YSG candidates is 0.05 mag in both the SMC and LMC. We therefore place any star that lies within 0.15 mag of the nearest single YSG model in this category, as it does not show evidence for an excess (relative to the models) at a level $\geq3\sigma$.
    \item \textit{Apparent blue/U$-$B excess observed:} Stars that are further from the single YSG models in U$-$B than the bounds described in 2) above. These are the candidate YSG+OB (cYSG+OB) binaries.
\end{enumerate}

The results of these cuts are shown in Table \ref{table_results}, and the classifications are plotted in V$-$I vs U$-$B space in the middle panels of Figure~\ref{fig:opticalcand}. Of the stars that fall into categories 2 and 3 (i.e.\ candidate YSGs that appear to robustly fall below our 9000 K temperature threshold), the fraction that have an apparent excess of blue light is 44\%/37\% in the SMC/LMC. These are the YSG+OB binary candidates. 

Example SEDs of two binary candidates with inferred YSG temperatures of $\sim$ 4000 and 6000 K T$_{\mathrm{eff}}$ are shown in Figure~\ref{fig:sed_example} alongside apparently single YSGs at the same temperatures. The magnitudes in these plots have been transformed to the AB system so that they are representative of the true spectral shape. The objects show similar slopes at infrared colors (hence our inference of similar temperatures for the YSGs), but clear differences can be seen at shorter wavelengths. In both cases, the stars we infer as apparently single display a significant drop off in flux in the UV, with UV magnitudes $\gtrapprox$ 5 mag fainter than the I-band. However, for the objects we identify as binary candidates this drop off is much less pronounced, with differences of only $\sim$2-3 mag.

We note that the fraction of YSGs with blue enough V$-$I colors that they fall in the first category above (i.e. with YSGs with T$_{\mathrm{eff}}>$9000 K according to the Pickles models) is higher in the LMC (40\%) than in the SMC (15\%). Upon examination, this may actually be due to the fact that we selected candidates YSGs from the same temperature range (4000 $<$T$_{\rm{eff}}$$<$9000K) for both galaxies. However, the Hayashi track is warmer in lower metallicity environments and is expected to fall near the low end of our adopted range in the SMC. Therefore, we may have included some stars from the base of the RSG branch in our SMC cuts, lowering the total fraction of stars at hot temperatures in the SMC (see e.g. the over-density of stars at low temperatures in the left panel of Figure~\ref{fig:hrd_ysgcand}).

\subsubsection{Optical Colors - Visual Method}\label{sec:optcol-visual}

We note that the classification method in the previous section relies on the assumption that the Pickles models accurately represent the observed spectral energy distributions of single YSGs. However, we highlight that we observe an apparent over-density of sources in a flat line at U$-$B = 1.3 (between V$-$I values of 1.25 and 3) in Figure \ref{fig:opticalcand}. This over-density appears at a slightly bluer U$-$B color than the theoretical models of single YSGs. These may truly be binary candidates, but it is also possible that the Pickles models may not reproduce the properties of observed YSGs precisely. 

\begin{figure}
    \centering
    \includegraphics[width=0.45\textwidth]{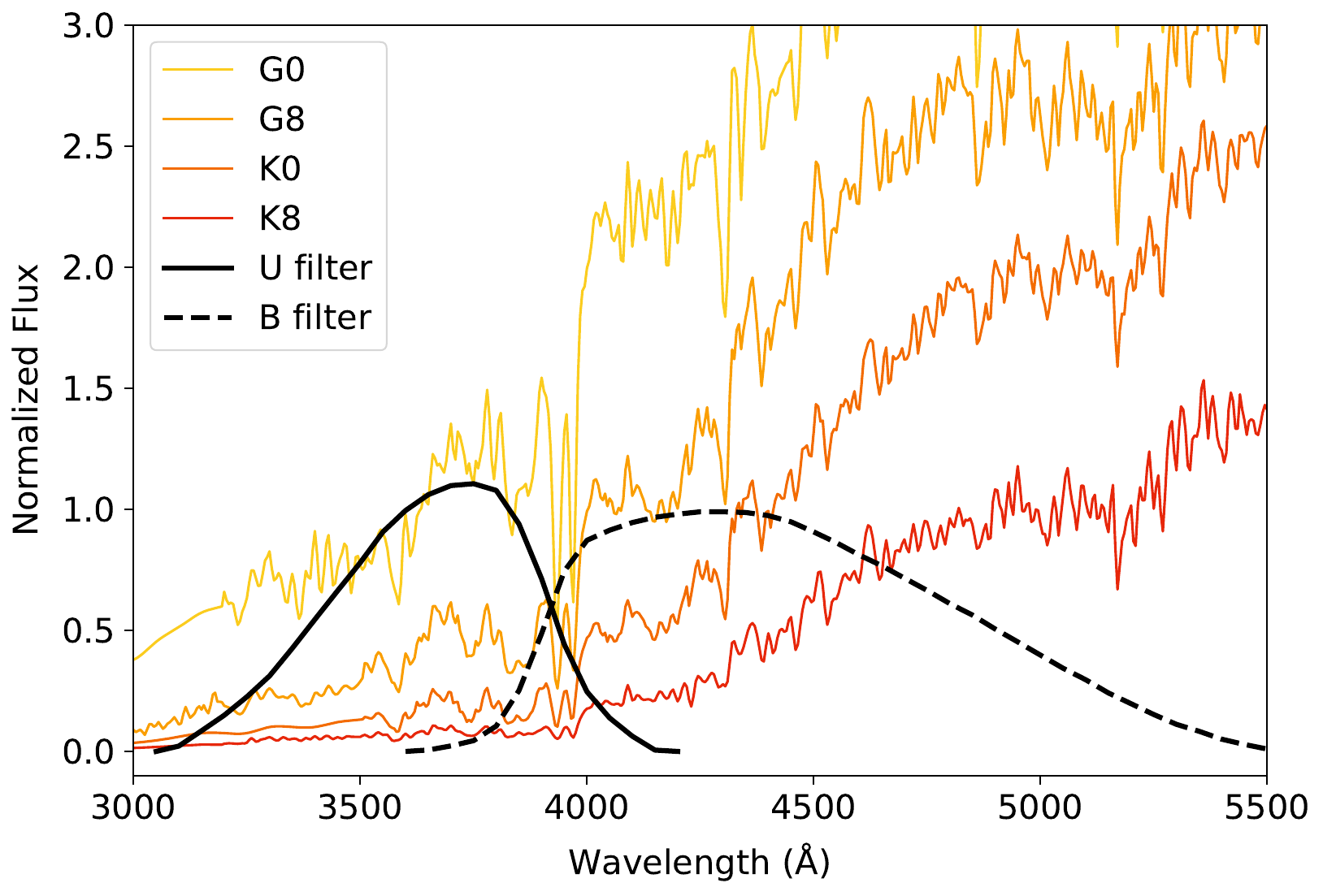}
    \caption{Pickles synthetic stellar spectra for G0, G8, K0, and K8 supergiants (orange to red lines) along with the filter profiles, arbitrarily scaled, for MCPS U- and B-band (black, solid and dashed lines respectively).}
    \label{fig:picklesubfilts}
\end{figure}

Importantly, the Pickles supergiant models are based on observations at solar metallicity, while the SMC and LMC have $\sim$1/5 and $\sim$1/2 solar metallicity, respectively \citep{Choudhury.S.2016.LMCMetallicity,Choudhury.S.2018.SMCMetallicity}. At short wavelengths (past the Balmer break at $\lessapprox$4000\AA), significant line blanketing greatly affects stellar spectral shape; see the examples in Figure~\ref{fig:picklesubfilts}. The magnitude of this line blanketing should be dependent on metallicity. Thus for lower Z environments such as the Clouds, we might expect a brighter U magnitude relative to the B magnitude---and therefore a smaller U$-$B color. This would have the effect of shifting both the single YSG and combined YSG+OB models down to lower U$-$B colors in Figures \ref{fig:combinedmodels} and \ref{fig:opticalcand}---potentially bringing the Pickles single YSG models closer to the center of a visual over-density of sources located around U$-$B = 1.3. 

\begin{figure*}
    \centering
    \includegraphics[width=0.9\textwidth]{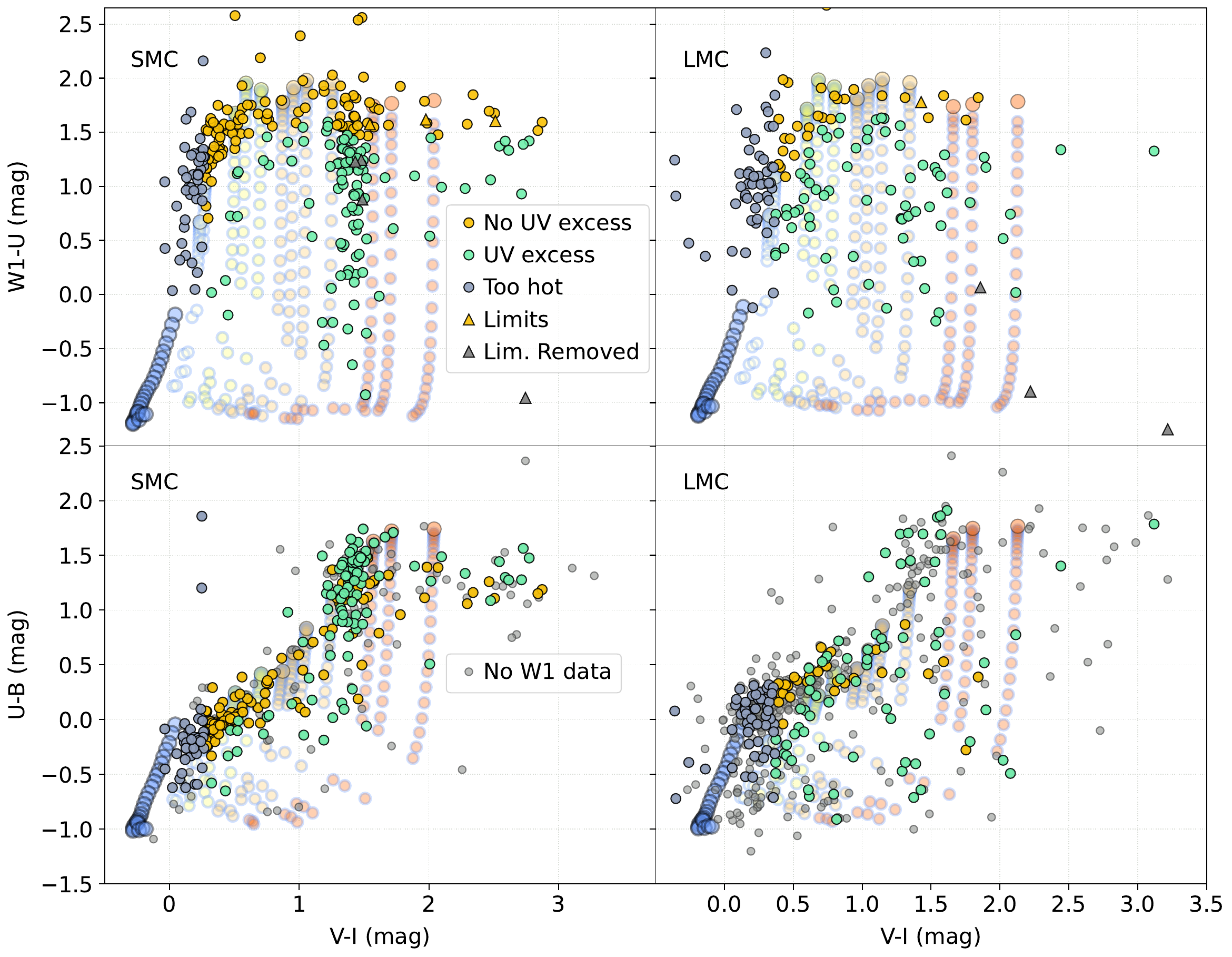}
    \caption{The same as Figure~\ref{fig:opticalcand}, but showing UV classifications. \textit{Top:} V$-$I vs W1$-$U color-color diagram. Limits are shown as triangles. \textit{Bottom:} The UV classifications from the top panels, but shown in V$-$I vs U$-$B space. The locations of the previous temperature cuts in U$-$B are shown with light grey lines. Stars without W1 data are show as faint grey circles.}
    \label{fig:uvcand}
\end{figure*}

While future work will examine the impact for stellar atmosphere models at a range of metallicities, it is possible that the observed over-density corresponds to the population of YSGs without a blue color excess. To account for this possible model-mismatch we now also perform a second `visual' (as in, by eye) inspection method of identifying stars with a blue excess. We first visually identify the areas of highest density in V$-$I vs U$-$B color space---using the models only as a general guide. We then designate all not inside the over-density (and with $V-I$ colors redder than the cut described in \S\ref{sec:optcol-model}) as having a blue excess. The results of this categorization can be seen in the bottom panels of Figure~\ref{fig:opticalcand}. When this method is applied the fraction of stars with blue excess drops to 16\%/26\% in the SMC/LMC (see Table \ref{table_results}).

\subsubsection{UV-Optical Colors}\label{sec:uvcol-results}

\begin{deluxetable}{l|rr}
	\centering
	\tablecaption{Photometric Classification Results\label{table_results}}
	\tablehead{\multicolumn{1}{c|}{Classification} & \colhead{SMC} & \colhead{LMC}}
	\startdata
    \multicolumn{3}{c}{Optical}\\\hline
    Total Stars & 382 & 481 \\
	Too hot YSGs & 59 & 191 \\
	Models: No U$-$B excess & 180 & 184 \\
    Models: U$-$B excess\tablenotemark{$\dagger$} & 143 & 106 \\
    Models: Blue excess fraction & 44\% & 37\% \\\hline
    Visual: No U$-$B excess & 271 & 214 \\
    Visual: U$-$B excess\tablenotemark{$\dagger$} & 52 & 76 \\ 
    Visual: Blue excess fraction & 16\% & 26\% \\\hline\hline
    \multicolumn{3}{c}{Ultraviolet}\\\hline
    Total Stars & 268* & 152* \\
	Too hot YSGs & 45 & 51 \\
	No W1$-$U excess & 118 & 30 \\
    W1$-$U excess\tablenotemark{$\dagger$}  & 105 & 71  \\
    UV excess fraction: & 41\% & 70\% \\
	\hline
	\hline
	\enddata
\tablenotetext{*}{Some upper limits removed, see text.}
\tablenotetext{\dagger}{Candidate YSG+OB systems.}
\end{deluxetable}
\vspace{-0.15in}

As described in Section~\ref{sec:criteria}, while optical colors are more readily available for stars in the clouds, incorporating UV photometry provides additional sensitivity to lower mass binary companions. We therefore repeat the same selection process using the UV-optical diagnostic diagrams, but with a few small differences. 

First, for the sample of candidate YSGs with W1 magnitudes available (see \S~\ref{sec:crossmatch}), the median W1$-$U uncertainty is 0.1 mag in the SMC and 0.05 mag in the LMC. To select a star as having a detectable ``UV excess'', we therefore require that they have W1$-$U colors that are 0.3 and 0.15 mag bluer than the nearest single YSG model, respectively. Second, we use only the model method for the UV color space, as there are not enough stars with UV data to confidently identify the `visual over-density' associated with likely single YSGs. Finally, if a star has an upper limit for it's W1 magnitude, we discard it from the analysis if its W1$-$U upper limit is within the `UV excess' area of color space, as we cannot discern if it truly has a W1$-$U excess or if the true W1 magnitude may be much fainter. This results in a lower number of `Total Stars' in the UV section of Table \ref{table_results} than Table \ref{ysg_selection}, 4 from both the SMC and LMC.

The results are given in Table \ref{table_results} and plotted in Figure~\ref{fig:uvcand} (top panels). From the stars with UV data available (which also have $V-I$ colors redder than the cut described in \S\ref{sec:optcol-model}) the fraction with a UV excess is 41\%/70\% in the SMC/LMC. The SMC fraction is much smaller than the LMC fraction, owing primarily to the higher W1$-$U uncertainty (the threshold to show an excess compared to the models at $\geq$3$\sigma$ is 0.3 mag, while in the LMC it is 0.15 mag). Due to this, we are less sensitive to certain mass binary companions in the SMC compared to what is plotted in Figure~\ref{fig:compmassteff}. If we take the smaller LMC uncertainty and apply it to the SMC data, the SMC fraction would rise to 64\%.

In order to compare our two methods, in Figure~\ref{fig:uvcand} (bottom panels), we plot the sources with W1 magnitudes available, color-coded by their UV categorizations on the \textit{optical} color-color diagram used in \S\ref{sec:optcol-model} and Figure~\ref{fig:opticalcand}. In total, there are 220/100 systems in the SMC/LMC have both optical and UV data available $V-I$ redder than the cutoff described in \S\ref{sec:optcol-model}. Of these, 50/34 have both a blue and UV color excess (as determined from the optical-only and optical-UV diagnostics, respectively), and are therefore strong binarity candidates. An additional 79/23 stars show no detectable excess in either color-color space---these are highly likely to be single YSGs or binaries with low mass companions below our detection thresholds (see Figure~\ref{fig:compmassteff}).

The remaining stars ($\sim$40\% in both galaxies with full UV/optical data and a $V-I$ below our threshold) were categorized differently between the two methods. 55/37 in the SMC/LMC showed a detectable UV excess in the UV-optical diagnostic diagram but no detectable blue excess in the optical-only diagnostic diagram. While detailed SED fitting is required, this may be indicative to YSG+B binaries with smaller companions than can be distinguished using our method and optical data alone. Finally, 36/6 stars in the SMC/LMC have an apparent blue color excess but do \emph{not} show a UV color excess. This is unexpected, as based on the models described in \S\ref{sec:combinedmodels} there is no area of parameter space where a blue flux excess is expected with no UV excess. This may represent data quality issues (particularly with the MCPS U$-$band) not caught by our SED vetting (\S~\ref{sec:sed-clean}) and also emphasizes the point made in \S\ref{sec:optcol-visual} that the Pickles atmosphere models may not fully represent the diversity of observed properties of YSGs and YSG binaries.

\section{Discussion}\label{sec:disc}

The goal of this work was to investigate a method of photometrically distinguishing YSG binary candidates in the Magellanic Clouds. We extended the optical photometry method of \citetalias{Neugent.K.2018.RSGBinaryMethod} to hotter temperatures, and added ultraviolet photometry. We identified over 100 YSG candidates with signs of blue or UV color excesses, possibly indicating the existence of OB-type companions. We will now discuss possible contaminants in the identified populations, the robustness of the method, and the implications of these preliminary results. A catalog of YSG+OB binary candidates will be included in a subsequent paper, after spectroscopic confirmation.

\subsection{Possible Contaminants}\label{sec:contaminants}

A full assessment of contaminants within, as well as the completeness of, our sample of candidate YSG+OB binaries requires spectroscopic confirmation of the nature of the stars. Nonetheless, we briefly discuss potential and known contaminants in our sample. It is important to consider both possible contaminants to our base YSG sample, as this will impact the denominator of any future assessment of YSG binary fractions, as well as specific types of sources that could masquerade as YSG+OB candidate systems based on our method.

\subsubsection{SIMBAD Cross match}

We cross matched the 400/481 YSG candidate stars in the SMC/LMC identified in \S\ref{sec:idYSG} that had good SEDs and complete MCPS data with the SIMBAD database \citep{SIMBAD.Wenger.M.2000}. 327/377 stars in the SMC/LMC have matches, and the primary SIMBAD classifications are shown in Table \ref{simbad}, as well as previously reported spectral types for 152/266 of the stars. Of the candidate YSG+OB binaries (as identified by the model method), 132/84 stars in the SMC/LMC have SIMBAD matches, including 50/49  with previous spectral classifications. 

The vast majority ($\sim$90\%) of YSG candidates with information in SIMBAD have been classified as stars, supergiants, or RSGs/YSGs. Some are classified as blue supergiants (BSGs). Examining our sample, these BSGs were all objects near the upper bound of the temperature range that we selected based on J$-$K colors (T$_{\mathrm{eff}}\sim$9000 K). Some of the stars in our YSG sample have listed spectral types in SIMBAD that are either O or B-type stars, both of which are typically associated with temperatures $>$10,000K. We first note that none of these stars with O- or B-type spectroscopic classification in SIMBAD are located in areas of ``no blue excess'' in Figure~\ref{fig:opticalcand}---they are all either binary candidates, or appear hotter than the A-type models. For some of the B-type stars, these may simply represent objects near the upper edge of our selected threshold, where J$-$K becomes much less sensitive to temperature and photometric uncertainty could plausibly admit a star with a T$_{\mathrm{eff}}$ up to 15,000 K (see \S\ref{init_cc}). Also, some of these stars may be YSG+OB binaries where the OB component was previously identified in a search for hot stars. Of particular interest, in SIMBAD 6 stars are classified as Be stars and 6 as Wolf-Rayet (WR) stars; we discuss these further below. 

If we focus on the sources we identified as YSG+OB binary candidates and have information in SIMBAD, 96\% of SMC candidates and 73\% of LMC candidates are also classified in SIMBAD as stars, supergiants, or RSG/YSGs. A small number are classified as BSGs or emission line stars, 1 SMC binary candidate is also a Be star candidate, and 4 LMC binary candidates are confirmed WR stars.

\begin{deluxetable}{l||cc|cc}
	\centering
	\tablecaption{SIMBAD Classifications\label{simbad}}
	\tablehead{\multicolumn{1}{c||}{Classifications} & \multicolumn{2}{c|}{cYSGs} & \multicolumn{2}{c}{cYSG+OB}\\
    \multicolumn{1}{c||}{} & \colhead{SMC} & \multicolumn{1}{c|}{LMC} &\colhead{SMC} & \colhead{LMC} }
	\startdata
	Total (Clean SED) & 400 & 481 & 143 & 106 \\ 
	SIMBAD Match & 327 & 377 & 132 & 84 \\\hline
    Star or SG & 106 & 263 & 26 & 43 \\
    RG/SG\tablenotemark{$\dagger$$\dagger$} & 149 & 18 & 79 & 7 \\
    YG/SG\tablenotemark{$\dagger$} & 51 & 39 & 22 & 11 \\
    BG/SG\tablenotemark{$\dagger$} & 3 & 15 & 2 & 7 \\
    Wolf-Rayet & 0 & 6 & 0 & 4 \\
    Emission Line & 5 & 13 & 2 & 7 \\
    Be Cand & 4 & 2 & 1 & 0 \\
    Other* & 9 & 21 & 0 & 5 \\\hline
    \multicolumn{5}{c}{Spectral Type (if available)}\\\hline
    O & 2 & 21 & 1 & 9 \\
    B & 9 & 54 & 2 & 19 \\
    A & 44 & 139 & 6 & 12 \\
    F & 22 & 25 & 4 & 3 \\
    G & 43 & 5 & 16 & 1 \\
    K/M & 32 & 22 & 21 & 5 \\
	\hline
	\hline
	\enddata
    \tablenotetext{\dagger}{G/SG = Giant/Supergiant}
    \tablenotetext{\dagger\dagger}{Includes Long Period Variables}
    \tablenotetext{*}{Includes variable (V), peculiar (PEC), carbon (C), and unidentified (**) stars.}
\end{deluxetable}

\subsubsection{Specific Contaminants: YSG sample}

We selected YSG candidates by identifying stars with infrared colors corresponding to temperatures between 4000 $<$ T$_{\mathrm{eff}} <$ 9000 K and luminosities above log(L/L$_{\odot}$) $\geq$ 4.0. This corresponds to YSGs with M$\geq$9M$_{\odot}$ from the Geneva evolutionary models \citep{Lejeune.T.2001.GenevaPhotModels}. Based on these Geneva models, the blue loops of lower mass stars do not intersect this area, thus we expect negligible contamination from lower mass yellow giants. However, lower mass stars in the post-AGB phase may transit through this area of the Hertzsprung-Russell Diagram on their way to becoming white dwarfs. We discuss post-AGB contamination further below. Finally, as mentioned in \S\ref{init_cc}, as the uncertainties on our temperature estimates above T$_{\mathrm{eff}}>$7000K are high, especially at the 9000K boundary, some very hot stars that may be considered blue supergiants instead of YSGs may have been included in the sample. We do not expect single MS/dwarf stars as a contaminant, as their K-band magnitudes would be far fainter than our cutoff.

\subsubsection{Specific Contaminants: YSG+OB Candidates}

We also consider stellar objects that may be mistaken for YSG+OB binaries. Specifically, while the goal of this paper was to identify stars with infrared colors consistent with YSGs but with extra blue light, we may have also picked up objects that are naturally bright in blue/UV, but have excess infrared light. Examples include post-AGB stars, Be stars and Wolf-Rayet stars. We discuss each of these cases, before also highlighting other possible contaminates such as single YSGs and chance line-of-sight alignments.

\emph{Post-AGB Stars:} While there are no known post-AGB stars in our YSG+OB binary candidate sample according to SIMBAD, we note that since a post-AGB star will contain both a hot stellar core \citep{Green.R.1986.postAGB,Jeffery.C.2010.postAGB} and dusty circumstellar material \citep{vanWinckel.H.2003.postAGB}, such sources may be a potential contaminant. To investigate this, we use MIST stellar evolution models \citep{Paxton.B.2011.MESAPaper0,Paxton.B.2013.MESAPaperI,Paxton.B.2015.MESAPaperII,Dotter.A.2016.MISTPaperI,Choi.J.2016.MISTPaperII} to calculate the lifetime of post-AGB stars that travel through the same region of J$-$K vs K color-magnitude space as our YSG selection criteria. We find that stars with masses $\geq$3M$_{\odot}$ pass through this region for 40-80 years. Using a method from \citet[\S6.1]{O'Grady.A.2020.superAGBidentification} and \citet[\S S6.4]{DroutGotberg.2023.IntStrippedStars} to estimate likelihood of post-AGB contamination by comparing the travel times above to the total population of SMC/LMC AGB stars and their total lifetimes. We find that the number of post-AGB stars expected in this region of color-magnitude space is $\lll$1.

\begin{figure*}
    \centering
    \includegraphics[width=0.9\textwidth]{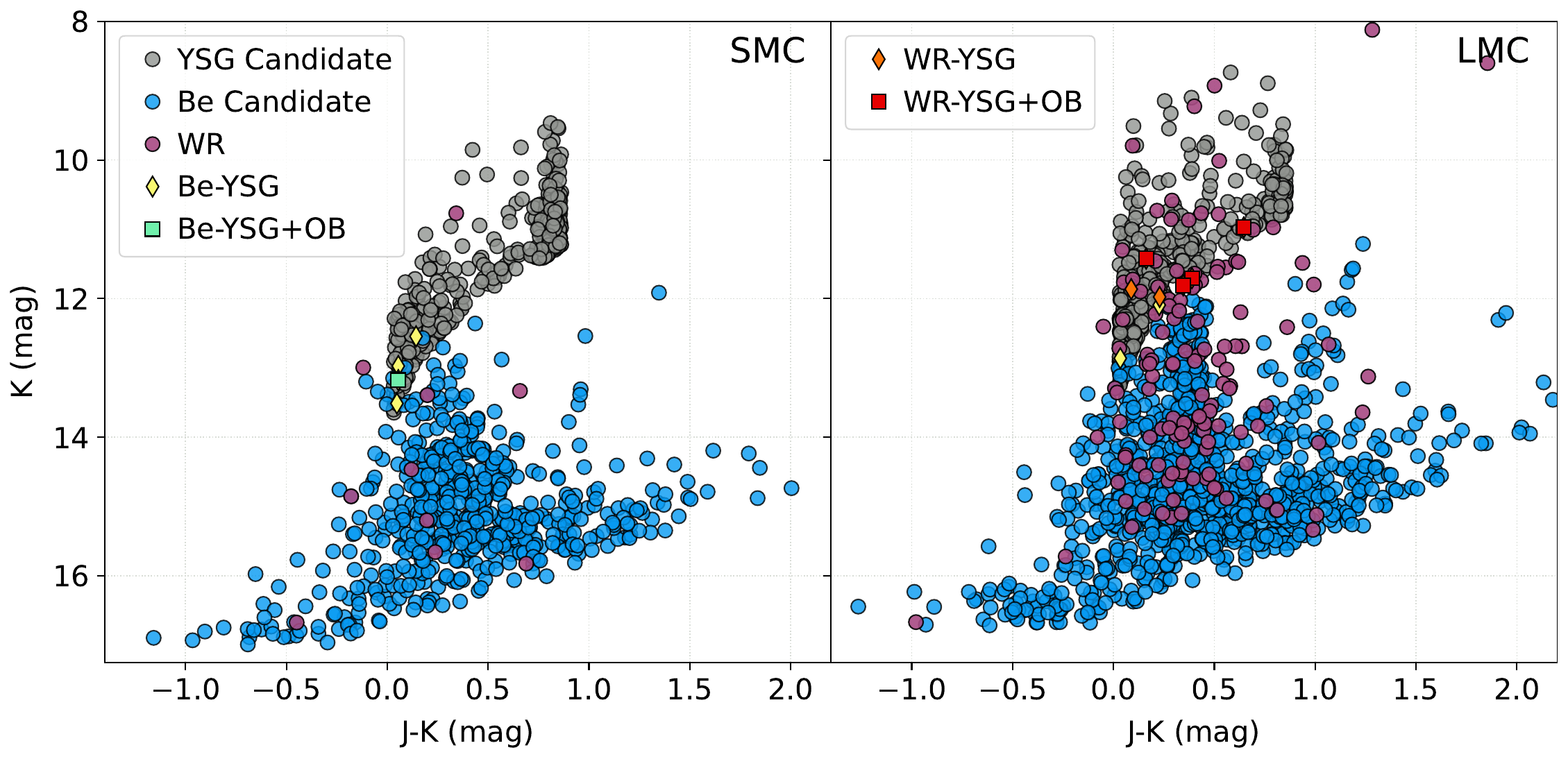}
    \caption{J$-$K vs K color-magnitude diagrams of the SMC (left) and LMC (right). Our YSG candidates are plotted as grey circles. The Be candidates of \citet[SMC]{Mennickent.R.2002.SMCBeStars} and \citet[LMC]{Sabogal.B.2005.LMCBeStars} are blue circles. Our YSG candidates that matched with Be stars are yellow diamonds, and YSG+OB candidates that matched are green squares. The census of SMC/LMC WR stars of \citet{Massey.P.2014.WRstarsMC} and \citet{Neugent.K.2018.WRCensus} are plotted in dark violet, and YSG or YSG+OB candidates that match are plotted as orange diamonds and red squares, respectively. We note that WR stars that overlap the YSG region but are not in our YSG sample were either removed as possible foreground contaminants or did not have complete MCPS data.}
    \label{fig:beplot}
\end{figure*}

\emph{Be stars:} Be stars are rapidly rotating B-type stars that form decretion disks that emit at infrared wavelengths. Therefore, Be stars share similar photometric profiles to YSG+OB binaries -- bright in both IR and UV wavelengths. We cross-matched our YSG candidates to the Magellanic Cloud Be catalogs of \citet[SMC]{Mennickent.R.2002.SMCBeStars} and \citet[LMC]{Sabogal.B.2005.LMCBeStars}, and found 4 matches in the SMC and 2 in the LMC. Our YSG candidates, the Be catalog populations, and these matches are plotted in Figure~\ref{fig:beplot}. All of these sources have extremely hot temperatures; all but one in the SMC have T$_{\mathrm{eff,V-I}}>$9000 K. As can be seen in Figure~\ref{fig:beplot}, the vast majority of Be stars are at dimmer K-band magnitudes than what our luminosity cut prescribed. Thus, while our YSG+OB star candidates may have some contamination from Be stars especially on the hot/faint end, given the expected faint magnitudes of most Be stars, and the wealth of emission line studies of the Clouds that have been completed to date \citep[e.g.,][]{MCELS.Smith.R.1999,Parker.Q.2005.SuperCosmos,Massey.P.2014.WRstarsMC}, we do not expect this to be a large source of contamination. Indeed, only 9 of the YSG+OB candidates in our sample are listed as emission-line objects in SIMBAD. 

However, we also note that \emph{YSG+Be} systems may be a possible example of an interacting binary system, where the Be star is a conservative mass gainer. \citet{Ramachandran.V.2023.PartiallyStrippedBinaryEx} recently identified a M$_{\mathrm{stripped}}\sim$3M$_{\odot}$ and M$_{\mathrm{Be}}\sim$17M$_{\odot}$ system, where the initial mass of the stripped star is predicted to have been $\sim$12M$_{\odot}$. The precursor of this system, just as mass transfer begins, could have been a $\sim$6000 K, log(L/L$_{\odot}$)$\sim$~4.6 YSG with an $\sim$11M$_{\odot}$ B star companion (see their Figure E.1). Thus we will carefully examine any confirmed Be stars in our sample for scenarios such as this.

\emph{Wolf-Rayet Stars:} 6 of the LMC YSG sample, of which 4 are YSG+OB candidates, are confirmed WR stars \citep{Henize.K.1956.WRLMC1,Fehrenbach.C.1976.WRLMC2,Melnick.J.1978.WRLMC3,Neugent.K.2018.WRCensus}. WR stars are incredibly bright in UV, and some cooler, carbon-oxygen dominated WR stars (also know by their WC designation) episodically form dusty, circumstellar shells that can be bright in infrared wavelengths \citep{Williams.P.1987.WRDusty,Crowther.P.2006.WRCensusWL1}. These shells come about due to binary interaction with a companion OB star \citep{Tuthill.P.1999.DustyWRStar,Crowther.P.2007.DustinWRs}. In the right panel of Figure~\ref{fig:beplot}, this overlap in IR colors can be seen. For our particular sample, the cataloging of WR stars in the Magellanic Clouds is likely complete \citep{Neugent.K.2018.WRCensus}, thus additional contamination would be unexpected.

Thus, while our final sample has some known contaminants from non-YSG+OB binary objects, we expect contamination from objects such as post-AGB stars, Be stars, and WR stars to be low. 

\emph{Single YSGs:} Instead, the dominant source of contamination is likely to be from single YSGs in the regions of parameter space where the level of blue/UV excess predicted is small, and thus any impact from uncertainties in our understanding of the UV flux from YSGs or dust extinction will be enhanced. As well, we note that some RSGs have shown significant near-UV excess when compared to standard stellar models. \citet{Massey.P.2005.DustRSGs} suggest this may be due to starlight being reprocessed by circumstellar dust around the star. Therefore, for YSGs that went through a windy RSG mass-loss phase, they may show a similar near-UV excess.

\emph{Line-of Sight-Alignments:} We also briefly discuss the possibility of chance alignment for our YSG+OB binary candidates, where a YSG and OB are not gravitationally bound but happen to `line up' from our perspective and appear as a binary when observed photometrically. \citet{Neugent.K.2020.RSGBinaryLMC}, while investigating RSG binaries in the LMC, ran Monte-Carlo simulations taking into account the OB star density around spectroscopically confirmed RSGs and found a 1.6\% $\pm$ 1.7\% chance that a RSG could have a line-of-sight companion. Given the good spatial agreement between RSGs and YSGs in the Clouds \citep[e.g.,][Fig. 5]{Neugent.K.2012.RSG.YSG.LMC} the same low percentage of chance alignment should apply to our YSG+OB binary sample, thus the chance of contamination from line-of-sight pairings is low.

\subsection{Validity of the Method}

We find that the method of \citetalias{Neugent.K.2018.RSGBinaryMethod} for identifying cool supergiants with B-type binary companions using optical photometry can in theory (e.g.\ based on theoretical model atmospheres) be successfully extended up to higher effective temperatures. The minimum mass of OB companion star that allows a YSG+OB binary to be discernible from a single YSG depends on the temperature and luminosity of the YSG (see Figure \ref{fig:compmassteff}). With optical photometry alone this method can discern YSG+OB binaries from single YSGs with OB star masses ranging from 3-6M$_{\odot}$ at 4000 K to 19-30M$_{\odot}$ at 9000 K. Adding UV photometry allows us to identify binary candidates with smaller companion masses, from 2-3M$_{\odot}$ at 4000 K and 7-16M$_{\odot}$ at 9000 K. 

One critical source of uncertainty in this method is the applicability of the models used to the full range of observed YSGs. As discussed in \S\ref{sec:optcol-visual}, the Pickles YSG models, particularly at cooler temperatures (G- and K-type) and in the SMC (lower metallicity), do not appear to align well in U$-$B with the observed population of YSG candidates (Figure~\ref{fig:opticalcand}, top left panel). This may stem from metallicity effects, since strong iron line blanketing decreases U-band magnitudes relatively more than B-band magnitudes. While the models are useful for identifying YSG+OB candidates, it is clear that existing YSG atmosphere models are not fully describing the observed properties of the YSG population in the Magellanic Clouds. Uncertainties in the true flux of the single YSG models have the greatest effect on stars that are ``close'' to the models, both YSG+OB binaries with low ($\lessapprox$5M$_{\odot}$) companion masses and truly single YSGs.

We also find few examples of contaminants in our final sample of YSG+OB candidates. Objects such as dusty Wolf-Rayet stars and Be stars are rare within the sample, and we similarly expect line-of-sight contamination to be low. The most likely source of contamination in the YSG+OB candidate sample would be from single YSGs erroneously identified as having a blue flux excess, either from uncertainty on photometry or extinction, or---for post-RSG YSGs---reprocessed starlight.

Therefore, we find that optical and UV photometry can be used to identify candidate YSG+OB binaries, with certain restrictions on YSG temperature and OB-star mass. However, in order to confirm what is causing the apparent excess of blue/UV light in these candidates, spectroscopic follow-up is essential, particularly for cases where the system may have a lower mass OB companion.

\subsection{Broad Implications of the fraction of stars with blue excess}

Without spectroscopic confirmation, we cannot make a robust estimate of the binary fraction of YSGs in the Magellanic Clouds. Additionally, as discussed above, the photometric method used is not sensitive to all possible YSG+OB binaries, as the discernibility of the binary system from a single YSG is dependent on the mass of the companion star and the YSG temperature and luminosity (see Figure \ref{fig:compmassteff}). With these caveats, we will briefly comment on the implications of our initial result for the fraction of stars with a blue excess to those without.

For the SMC/LMC, with optical photometry we find a blue excess fraction of 44\%/37\% when comparing photometry directly to models, and 16\%/26\% when using the visual method. \citet{Neugent.K.2021.RSGBinM3133} found the binary fraction within M31 to increase with increasing metallicity; here the visual method results follow this trend. The fraction of stars with a UV excess is 41/70\%, though the SMC fraction rises to 64\% if the photometric uncertainty is lowered to LMC levels, and these percentages are dependent on which stars have detectable UV flux. 

These fractions are all below the 70-90\% binary fraction for massive stars on the MS \citep{Sana.H.2012.BinaryFraction,Kobulnicky.H.2014.BinariesOB2,Offner.S.2023.MultipleStarSystems}, but generally above the 20\% observed for RSGs in the LMC. This RSG binary fraction measured by \citet{Patrick.L.2017.RSGBinNGC55,Patrick.L.2019.RSGBin30Dor,Patrick.L.2020.RSGBinNGC330,Neugent.K.2020.RSGBinaryLMC,Neugent.K.2021.RSGBinM3133} is likely dominated by wide binaries that have not yet interacted. In contrast, it is thought that up to 20-30\% of main sequence binaries will merge and another 40-50\% will undergo mass stripping via RLOF \citep{Sana.H.2012.BinaryFraction}. We might then expect the binary fraction of YSGs to be higher than 20\% due to binaries that will interact or that are in the process of interacting in the Hertzsprung Gap and that will not reach the RSG branch.

\subsection{Notable Regions of the Parameter Space}

Some of the YSG+OB binary candidates identified in this paper have extremely large blue color excesses, e.g., the stars that align with combined models with OB-star masses $>$20M$_{\odot}$ in Figure~\ref{fig:opticalcand}. 15 stars in the SMC and 58 in the LMC lie in this area. Some of these systems may simply be non-interacting YSG+O binaries. \citet{Neugent.K.2020.RSGBinaryLMC} find in a BPASS \citep{Eldridge.J.2017.Models2Binaries} simulation of RSG binaries that $\sim$1\% have O-type companions, so YSG+O binaries would similarly be expected to be rare, but not impossible. However, the stars we see in this area represent $\sim$10\%/55\% (SMC/LMC) of the total YSG+OB candidate population. It is therefore unlikely that these are all non-interacting YSG+O binaries.

Some of these systems with extremely strong blue color excesses may also be post-interaction systems. As discussed in \S\ref{sec:companionsofysgs}, multiple populations of stars with varied evolutionary histories are predicted to exist in the Hertzsprung Gap. This includes RLOF mass transfer systems (Case B mass transfer) that will lead to  partially stripped YSGs, the progenitors of Type IIb SNe \citep{Yoon.Sung-Chul.2017.TypeIbIIbBinarity,Zapartas.E.2017.SESNeCompanions,Klencki.J.2022.PartialEnvelopeMissingStars}, and systems where the stripped primary star expands and re-crosses the Hertzsprung Gap \citep{Laplace.E.2020.SEStars}. For the case of conservative mass transfer, the secondaries of these systems should be massive B- or O-type stars, and in some cases will now be more massive than the primary. Therefore, these objects may be particularly promising candidates for post-interaction binary systems. 

\subsection{Future Work: Towards a YSG Binary Fraction and Identification of Binary Interaction Products}

Here, we briefly outline the future direction of this work, which will be presented in a series of upcoming papers. First, using our ongoing observational programs, we will investigate the true nature of YSG+OB binary candidates presented here by searching for spectroscopic signatures of a binary companion (using similar methods to e.g. \citealt{Neugent.K.2019.RSGBinaryDiscovering}). Leveraging these results, we will compute the binary fraction for YSGs in the Magellanic Clouds, which---when coupled with results from other evolutionary phases \citep[e.g.][]{Sana.H.2012.BinaryFraction,Patrick.L.2019.RSGBin30Dor,Neugent.K.2020.RSGBinaryLMC}---can provide a critical benchmark for binary population synthesis models. Subsequently, we will search of individual systems that will have had their evolutionary pathways altered by binary interaction. While the evolutionary state of a YSG cannot be determined by a single measurement of its temperature and luminosity, differences in other properties are predicted. We highlight a number of these properties here. In practice, it will likely be necessary to combine multiple methods to place detailed constraints on specific objects. 
\emph{Orbital Solutions:} First, constraining the orbital solutions of confirmed YSG binaries can provide additional context on which systems that they may interact in the future. Since the radii of YSGs are large ($\sim10^{2}$--$10^{3}$ R$_{\odot}$) the periods of the systems are expected to be long ($\gtrsim$ 200 days) and the amplitude of any radial velocity variations moderate ($\lesssim$50 km s$^{-1}$). However, we note that i) these stars are bright, so smaller telescopes could be used, and ii) the low surface gravities of the YSGs will lead to very narrow lines, thus small RV deviations should be more easily detectable. In addition, a wealth of high cadence, long baseline time-domain surveys include the Magellanic Clouds (e.g. OGLE \citealt{Udalski.A.2003.OGLE,Ulaczyk.K.2013.OGLELMCVars}, ASAS-SN \citealt{Shappee.B.2014.ASASSN1,Kochanek.C.2017.ASASSN2,Pawlak.M.2019.ASASSNIV,Jayasinghe.T.2019.ASASSNVI,Jayasinghe.T.2019.ASASSNVII,Jayasinghe.T.2020.ASASSNV}), and we can use these datasets to search for evidence of binarity in the YSG+OB candidate light curves. In particular: (i) for certain inclinations, eclipses may be visible in the light curves and (ii) some of the YSG binary systems may show ellipsoidal variation in their light curves. While ellipsoidal variation is typically observed for much tighter binary systems, because the envelopes of YSGs are so sparse they can be more easily deformed.

\emph{Abundances:} The relative abundances of H and He, as well as C, N, and O are expected to be modified in YSGs that have been stripped (either though binary interaction or due to RSG winds). For example, \citet{Laplace.E.2020.SEStars} recently performed calculation for helium giant stars. These are objects that are fully (or nearly fully) stripped of hydrogen and lived on the He main sequence during He core burning (e.g. those identified by \citealt{DroutGotberg.2023.IntStrippedStars}) but subsequently re-expand and cool after core helium-burning has completed. \citet{Laplace.E.2020.SEStars} find that these objects should show a dearth of H and and enhancement in both He and N on their surfaces. In addition, \citep{Saio.Hideyuki.2013.PulsationYSGStrange} find that YSGs which have previous been through a RSG phase should have a 
N/C ratio $\sim$20\% higher than a YSG that hasn't. Recently, \citet{Ramachandran.V.2023.PartiallyStrippedBinaryEx,Ramachandran.V.2024.StrippedStarII} identified several partially stripped+Oe/Be binary systems hidden within the main sequence (thus at hotter temperature than our search) through observing an extreme CNO abundance pattern---demonstrating the utility of these constraints if they can be obtained.

\emph{Luminosity-to-Mass Ratio:} YSGs that have undergone stripping are also expected to have different luminosity-to-mass ratios when compared to YSGs on their first passage across the HRD. This is especially true for inflated helium giants, for which \citet{Laplace.E.2020.SEStars} found extremely low surface gravities (log(g)$\approx$-0.4). It is possible that the different luminosity-to-mass ratio of stripped YSGs many manifest in their pulsation properties \citep{Saio.Hideyuki.2013.PulsationYSGStrange}.

\emph{CSM Environment:} Finally, the circumstellar environment of currently interacting or post-interacting binaries may be distinct. Through spectroscopy we can also search for signatures of current accretion and/or recent mass loss from the system (e.g. H$\alpha$ emission, infrared excess, nebulae features in the 2D spectrum).

\section{Summary}\label{sec:concl}

\emph{Motivation:} Our goal of this paper was to determine a method to photometrically separate single YSG stars from YSG binaries and then apply that method to observations of stars in the Small and Large Magellanic Clouds to identify a set of candidate YSG+OB binary systems. To do this, we constructed models of YSG+OB binaries by combining stellar atmosphere models of YSGs and OB-type MS stars. We then compared the optical and UV photometric colors of these models to those of hundreds of YSG candidates in the Magellanic Clouds.

\emph{Results:} We found, depending on the method, $\sim$150-250 YSG candidates that showed evidence of blue or UV color excesses, a possible sign of OB-type companions. In general, optical photometry can be used to identify YSG+OB binary candidates with companion masses as low as 3-6M$_{\odot}$ (for YSG log(L/L$_{\odot}$) = 4.2 - 5.1) at low ($\sim$4000 K) temperatures, but the minimum companion mass required to discern a binary from a single YSG rises to 19-30M$_{\odot}$ at higher temperatures ($\sim$9000 K). At these hotter temperatures, UV photometry can probe companion masses down to 7-16M$_{\odot}$. Binaries with very large O-star masses can be identified at all temperatures. 

We find that we expect only minimum contamination in our photometric candidates from other classes of objects such as Be stars and dusty Wolf-Rayet stars. However, we find that some models of YSGs may not accurately reflect the observed colors of YSGs, especially at shorter wavelengths. As a result, some contamination from single YSGs may be possible, especially for objects with smaller amounts of observed UV/blue excess (e.g. low companion masses). This emphasizes the need for spectroscopic follow-up of our sample. 

We report a fraction of stars with a discernible blue excess of 44\% in the SMC and 37\% in the LMC, when directly comparing to stellar atmosphere models. When we instead perform a visual categorization (see \S\ref{sec:optcol-visual}), the fractions are 16\% in the SMC and 26\% in the LMC.

\begin{acknowledgements}

\section*{Acknowledgments}

The authors thank: Aaron Tohuvavohu, Katie Breivik, Marten van Kerkwijk, Jakub Klencki, Eva Laplace, and Dae-Sik Moon for helpful discussions, and Adiv Paradise for helpful edits. The authors also thank the anonymous reviewer for a helpful and constructive referee report. 

The authors at the University of Toronto acknowledge that the land on which the University of Toronto operates is the traditional territory of the Huron-Wendat, the Seneca, and the Mississaugas of the Credit River. They are grateful to have the opportunity to work on this land.

The Dunlap Institute is funded through an endowment established by the David Dunlap family and the University of Toronto.

A.J.G.O. is supported by a McWilliams Fellowship at Carnegie Mellon University. M.R.D. acknowledges support from the NSERC through grant RGPIN-2019-06186, the Canada Research Chairs Program, and the Dunlap Institute at the University of Toronto. B.M.G. acknowledges the support of the Natural Sciences and Engineering Research Council of Canada (NSERC) through grant RGPIN-2022-03163, and of the Canada Research Chairs program. Support for this work was provided by NASA through the NASA Hubble Fellowship Program grant Nos. HST-HF2-51457.001-A and HST-HF2-51516 awarded by the Space Telescope Science Institute, which is operated by the Association of Universities for Research in Astronomy, Inc., for NASA, under contract NAS5-26555.

This research has made use of the SIMBAD database \citep{SIMBAD.Wenger.M.2000}, operated at CDS, Strasbourg, France, and the SVO Filter Profile Service (\url{http://svo2.cab.inta-csic.es/theory/fps/}) supported from the Spanish MINECO through grant AYA2017-84089 \citep{Rodrigo.C.2012.SVOFilterService}. This research has made use of the following software: \software{astropy \citep{Astropy.2013.AstropyI,Astropy.Collab.2018.Astropy,Astropy.2022.AstropyIII}, IRAF \citep{Tody.D.1986.IRAFI,Tody.D.1993.IRAFII}, TOPCAT \citep{Taylor.M.2005.TOPCAT}.}

\end{acknowledgements}

\bibliography{thesis_bib}
\bibliographystyle{aasjournal}

\end{CJK*}
\end{document}